\documentclass[aps,pre,reprint]{revtex4-1}

\usepackage[utf8]{inputenc}
\usepackage{amsmath}
\usepackage{amsfonts}
\usepackage{amssymb}
\usepackage{amsthm}
\usepackage{amsopn}
\usepackage{upgreek}
\usepackage{bm}
\usepackage{epsfig}
\usepackage{nicefrac}

\renewcommand{\eqref}[1]{Eq.~\ref{#1}}
\newcommand{\eqsref}[1]{Eqs.~\ref{#1}}
\newcommand{\figref}[1]{Figure~\ref{#1}}
\newcommand{\figsref}[1]{Figures~\ref{#1}}
\newcommand{\smfigref}[1]{Supplementary Methods Figure~\ref{#1}}
\newcommand{\tabref}[1]{Table~\ref{#1}}

\newcommand{\secref}[1]{Sec.~\ref{#1}}

\newcommand{\refcite}[1]{Ref.~\onlinecite{#1}}

\newcommand{\kB}{k_{\text{B}}}
\newcommand{\Ecoli}{\textit{E.~coli}}

\begin{document}

\newcommand{\addressharvard}{Department of Chemistry and Chemical Biology, Harvard University, 12 Oxford Street, Cambridge, MA, 02138, USA}

\title{Evidence of evolutionary selection for co-translational folding}
\author{William M.~Jacobs}
\affiliation{\addressharvard}
\author{Eugene I.~Shakhnovich}
\affiliation{\addressharvard}
\date{\today}

\begin{abstract}
Recent experiments and simulations have demonstrated that proteins can fold on the ribosome.
However, the extent and generality of fitness effects resulting from co-translational folding remain open questions.
Here we report a genome-wide analysis that uncovers evidence of evolutionary selection for co-translational folding.
We describe a robust statistical approach to identify loci within genes that are both significantly enriched in slowly translated codons and evolutionarily conserved.
Surprisingly, we find that domain boundaries can explain only a small fraction of these conserved loci.
Instead, we propose that regions enriched in slowly translated codons are associated with co-translational folding intermediates, which may be smaller than a single domain.
We show that the intermediates predicted by a native-centric model of co-translational folding account for the majority of these loci across more than 500 \Ecoli{} proteins.
By making a direct connection to protein folding, this analysis provides strong evidence that many synonymous substitutions have been selected to optimize translation rates at specific locations within genes.
More generally, our results indicate that kinetics, and not just thermodynamics, can significantly alter the efficiency of self-assembly in a biological context.
\end{abstract}

\maketitle

\section*{Introduction}

Many proteins can begin folding to their native states before their synthesis is complete~\cite{komar2009pause,pechmann2013ribosome}.
As much as one-third of a bacterial proteome is believed to fold co-translationally~\cite{ciryam2013vivo}, with an even higher percentage likely in more slowly translated eukaryotic proteomes.
Numerous experiments on both natural and engineered amino-acid sequences have shown that folding during synthesis can have profound effects: compared to denatured and refolded chains, co-translationally folded proteins may be less prone to misfolding~\cite{netzer1997recombination,frydman1999cotranslational,komar1999synonymous,kim2015translational,siller2010slowing,ugrinov2010cotranslational,agashe2013good,clark2001newly}, aggregation~\cite{evans2008cotranslational} and degradation~\cite{zhang2009transient}, or they may preferentially adopt alternate stable structures~\cite{sander2014expanding,buhr2016synonymous,zhou2013non}.
Because the timescales for protein synthesis and folding are often similar~\cite{obrien2014understanding,nissley2016accurate}, it is clear that the rate of translation can be used to tune the self-assembly of peptide chains \textit{in vivo}~\cite{xu2013non,kimchi2007silent}.
To this point, however, there exists little evidence that evolution has selected specifically for efficient co-translational folding kinetics across any substantial fraction of an organism's proteome.

In this work, we provide evidence that evolutionary selection has tuned protein-translation rates to optimize co-translational folding pathways.
Our approach is motivated by the hypothesis that pauses during protein synthesis may be beneficial for promoting the formation of native structure.
By increasing the separation between the timescales for folding and translation, such pauses may promote the assembly of on-pathway intermediates, which, in turn, template the growth of further native structure.
Many experimental and computational studies have shown that protein-folding naturally proceeds in a step-wise manner via structurally distinct intermediates~\cite{hartl2009converging,jacobs2016structure}, and that cooperative folding cannot commence until a minimal number of residues have emerged from the ribosome exit tunnel~\cite{eichmann2010cotranslational,holtkamp2015cotranslational,elcock2006molecular,obrien2012prediction}.
These~general~findings suggest that any beneficial pauses during synthesis should occur at specific locations within an amino-acid sequence.

Using a coarse-grained model of co-translational folding, we find that translational pauses tend to be associated with stable, native-like co-translational folding intermediates.
The relevant folding intermediates are typically not complete structural domains, as has often been assumed~\cite{jacobson2016quality}, and may be distinct from intermediates that are observed when refolding from a denatured ensemble.
By comparing putative translational pause sites with a neutral model that accounts for gene-specific codon usage, we show that evolutionarily optimized co-translational folding is a widespread feature of the \Ecoli{} genome.
Our results therefore highlight the extent to which evolution has tuned the self-assembly pathways, and not just the native structures, of complex biomolecules.

\begin{figure*}
  \includegraphics[width=0.9\linewidth]{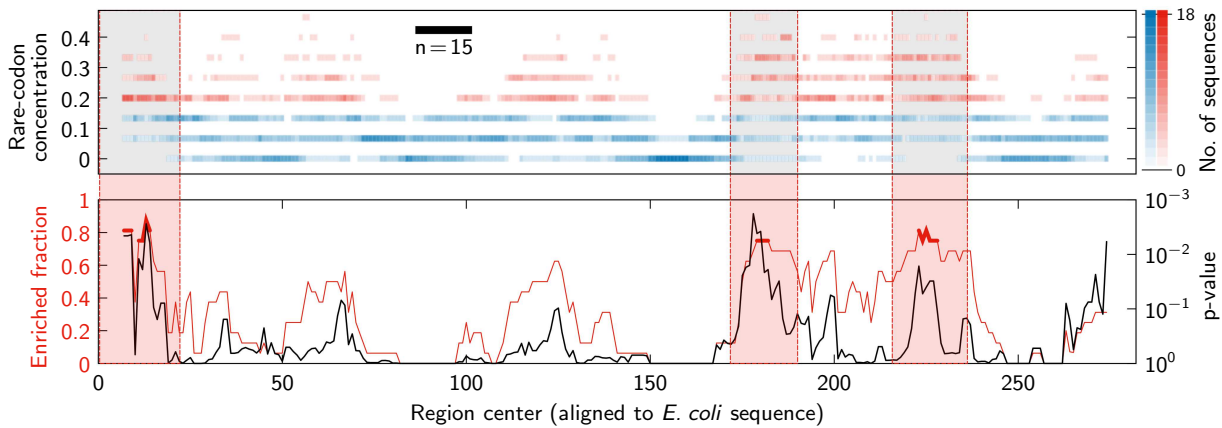}
  \caption{An example multiple-sequence alignment identifies conserved rare-codon enrichment within the gene \textit{folP}.  Above, a histogram shows the number of sequences that have a given local concentration of rare codons at each position in the alignment.  The local concentrations of rare codons are determined within 15-codon regions.  Based on the average occurrence of rare codons in \textit{folP}, local rare-codon concentrations of at least 3/15 are considered to be enriched and are colored red in the histogram, while 15-codon regions with fewer than three rare codons are not enriched and are shown in blue.  Below, the fraction of sequences that are enriched at each position in the alignment (red) is shown along with the corresponding neutral-model p-value (black), as explained in the main text.  Conserved regions, where at least 75\% of the sequences are enriched, are highlighted.}
  \label{fig:msa}
\end{figure*}

\section*{Results}

\subsection*{Unbiased identification of slowly translated regions}

Our analysis of beneficial pauses in protein synthesis relies on the identification of regions within mRNA transcripts that are enriched in `rare' codons (SI Appendix, Table~S1), i.e.~codons that are used substantially less often than alternate synonymous codons in highly expressed genes~\cite{sharp1987codon}.
Despite numerous attempts to predict codon-specific translation rates based on physical factors~\cite{dana2014effect,li2012anti,pelechano2015widespread,weinberg2016improved}, such as tRNA concentrations, translation-speed estimates based on relative-usage metrics~\cite{sharp1987codon,clarke2008rare} remain among the most accurate~\cite{yu2015codon,chaney2015roles,spencer2012silent}.
Thus, using codon rarity as a proxy for translation speed, we can look for pauses in synthesis by identifying regions in a mRNA transcript that are locally enriched in rare codons.

However, an appropriate neutral model must account for two potential sources of synonymous codon-usage bias at the level of an individual gene.
First, we controlled for the overall rare-codon usage in a gene, which is defined as the fraction of rare codons in the entire transcript (SI Appendix, Fig.~S1).
Multiple factors have been hypothesized to contribute to the overall degree of codon adaptation of each gene, including evolutionary selection for rapid synthesis, accurate translation and the stability of mRNA transcripts~\cite{plotkin2011synonymous}.
By taking a gene's average codon usage into account, we instead pick out regions that are locally enriched in rare codons relative to the gene-specific background.
Second, we accounted for synonymous-codon bias due to the amino-acid composition of the protein sequence.
Assuming that amino-acid sequences are under stronger selection pressure and can thus be considered immutable, we estimated the average rare-codon frequencies for each amino-acid type among all genes with a similar level of rare-codon usage.
Having controlled for the overall rare-codon usage and the amino-acid sequence, we modeled neutral codon usage as a Bernoulli process with sequence-dependent rare-codon probabilities (see SI Appendix, Sec.~1A).

\subsection*{Evaluation of evolutionary conservation}

Next, we assessed the functional importance of local rare-codon enrichment by looking for conservation of rare-codon usage across multiple-sequence alignments (\figref{fig:msa}).
We extended the neutral model described above to 18 sufficiently diverged prokaryotic genomes, with rare-codon definitions and gene-specific rare-codon probabilities computed for each genome independently.
Here our approach differs from conventional conservation analyses, because we are interested in the enrichment of rare codons within contiguous 15-codon segments of a transcript, as opposed to the codon usage at each aligned site~\cite{pechmann2013evolutionary,widmann2008analysis}.
By examining conservation of rare-codon enrichment, we can identify local regions that do not align precisely but nevertheless result in translational pauses at similar places within the protein sequence.
This approach also allows for a meaningful comparison of the local rare-codon enrichment in sequence alignments that contain insertions and deletions.
Our choice of a 15-codon enrichment region is comparable to the length of a typical element of protein secondary structure, and we verified that regions with widths of 10 and 20 codons yield similar results.
In contrast, larger enrichment regions defined on the basis of complete domains rarely differ significantly from the background rare-codon usage, while analyses of single aligned sites tend not to produce statistically significant results.

\begin{figure*}
  \includegraphics[width=0.9\linewidth]{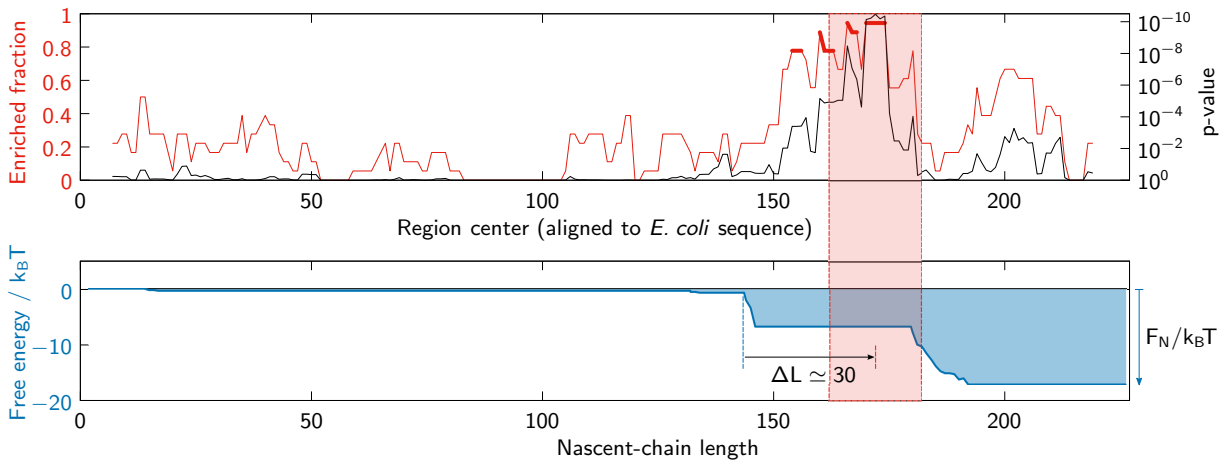}
  \caption{Predicted co-translational folding intermediates correspond to highly conserved regions of rare-codon enrichment.  Above, the fraction of enriched sequences and corresponding p-values for the gene \textit{cmk} are shown as in \figref{fig:msa}.  Below, the minimum free energy, relative to the unfolded ensemble, of a nascent chain of length $L$ is shown by the solid blue line; the stability of the native full-length protein is ${F_N / \kB T}$.  Native-like intermediates become stable where this minimum-free-energy curve decreases sharply.  The lowest p-value for enriched sequences (highlighted region) is approximately ${\Delta L = 30}$ codons downstream of the first predicted folding intermediate.}
  \label{fig:cotranslational_folding}
\end{figure*}

To be relevant for co-translational folding, putative slowly translated regions must meet two criteria: a high degree of conservation of slowly translated codons, and a low probability of such an occurrence in the neutral model.
For a region to be considered both enriched and conserved, we required that the local concentration of rare codons deviate from the background distribution by approximately one standard deviation in at least 75\% of the sequences in the alignment; Fig.~S2 of the SI~Appendix shows that our results are robust with respect to this conservation threshold.
We then computed an associated p-value that reports the probability, within the neutral model, of randomly generating at least the observed number of enriched regions from reverse translations, i.e., by sampling synonymous sequences using the aligned amino-acid sequences and a probabilistic model of the codon usage for each amino-acid type (see SI Appendix, Sec.~1A).
This second criterion is central to our findings, as we shall discuss below.
We emphasize that these criteria are distinct: depending on the amino-acid identities, it is possible to observe low p-values without significant rare-codon enrichment relative to the background, and vice versa.
Consequently, both criteria must be satisfied to constitute evidence for evolutionary selection.

Our analysis reveals numerous rare-codon enrichment loci in the \Ecoli{} genome that are inconsistent with the neutral model, and are thus likely to be a result of evolutionary selection (SI Appendix, Fig.~S3a).
Although these regions occur throughout the mRNA transcripts, their locations are biased towards both the 5' and 3'-ends (SI Appendix, Fig.~S3b).
While these trends have been noted previously~\cite{clarke2010increased}, our analysis confirms that the increased probability of rare-codon enrichment at the 3'-end is evolutionarily conserved and is not a consequence of the amino-acid sequences.
Furthermore, we find that these biases become more pronounced as we lower the p-value threshold used for comparison with the neutral model (SI Appendix, Fig.~S3b), suggesting that any false positives from our analysis are relatively evenly distributed throughout the transcripts.
We also analyzed the codon-level similarity among the genomes in our alignments and verified that these results reflect conservation of rarity as opposed to conservation of specific codons (SI Appendix, Fig.~S4).

\subsection*{Comparison with predicted co-translational folding pathways}

To probe the potential consequences of local rare-codon enrichment for protein folding, we next examined the formation of native-like intermediates during protein synthesis.
A large body of simulation evidence~\cite{trovato2016insights} has shown that intermediates must be stable at equilibrium in order to be sampled with high probability during co-translational folding and are only likely to form when the folding rate is fast relative to the protein elongation rate.
Therefore, while an intermediate's equilibrium free energy does not completely determine whether it will appear on a co-translational folding pathway, we assume that stability at equilibrium is necessary for a pause in translation to promote the development of native structure.

Here we applied a coarse-grained model~\cite{jacobs2016structure} to predict the formation of stable partial structures during nascent-chain elongation.
Importantly, this model captures the tertiary structure of nascent chains and does not assume that domains fold cooperatively or independently.
To model co-translational folding, a nascent chain of length $L$ is allowed to form native contacts among the first $L$ residues of the full protein.
We then computed the minimum free energy of a nascent chain, relative to an unfolded ensemble, using a mean-field theory based on the protein's native structure (see SI Appendix, Sec.~1B).
This approach captures the opposing contributions to the free energy from energetically favorable native contacts and the configurational entropy of an unfolded chain.
We used a native-centric energy function that emphasizes hydrogen bonds and contacts between larger residues~\cite{jacobs2016structure}, while the thermodynamic stability of the native state is fixed based on the full protein length~\cite{ghosh2009computing}.
We show in Fig.~S5 of the SI~Appendix that tuning the native-state stability does not significantly affect the results of our analysis.

Our calculations predict that, in general, native structure forms discontinuously during nascent-chain elongation.
In the example shown in the lower panel of \figref{fig:cotranslational_folding}, decreases in the nascent-chain free energy occur at distinct chain lengths.
These sudden drops correspond to the appearance of stable intermediates with native-like tertiary structure.
In contrast, at chain lengths corresponding to the intervening plateaus, the nascent-chain free energy remains constant because the newly synthesized residues cannot form sufficient stabilizing contacts with any existing tertiary structure.
Unsurprisingly, the probability of finding a stable on-pathway intermediate increases as synthesis nears completion (SI Appendix, Fig.~S6).

We are now in a position to test the relationship between translational pausing and the formation of native-like intermediates.
The ribosome exit tunnel is widely believed to conceal between 30 and 40 amino acids~\cite{gloge2014co,chaney2015roles}, although a greater number may be accommodated in partially helical conformations~\cite{lu2005folding}.
In addition, some tertiary structure formation may commence within the exit-tunnel vestibule~\cite{nilsson2015cotranslational}.
A beneficial pause in synthesis should therefore be separated from a co-translational intermediate by a distance that is roughly equivalent to the exit-tunnel length (see Methods).
An example of this correspondence is shown in \figref{fig:cotranslational_folding}, where a putative translational pause is located approximately 30 residues downstream of the formation of a predicted intermediate.
However, we emphasize that, according to the present hypothesis, the formation of an intermediate is necessary but not sufficient to expect that a translational pause would be beneficial.
For example, intermediates that fold quickly relative to the average translation rate or appear less than the exit-tunnel distance from the end of the protein are unlikely to be accompanied by a productive pause.

\begin{figure}
  \includegraphics[width=\linewidth]{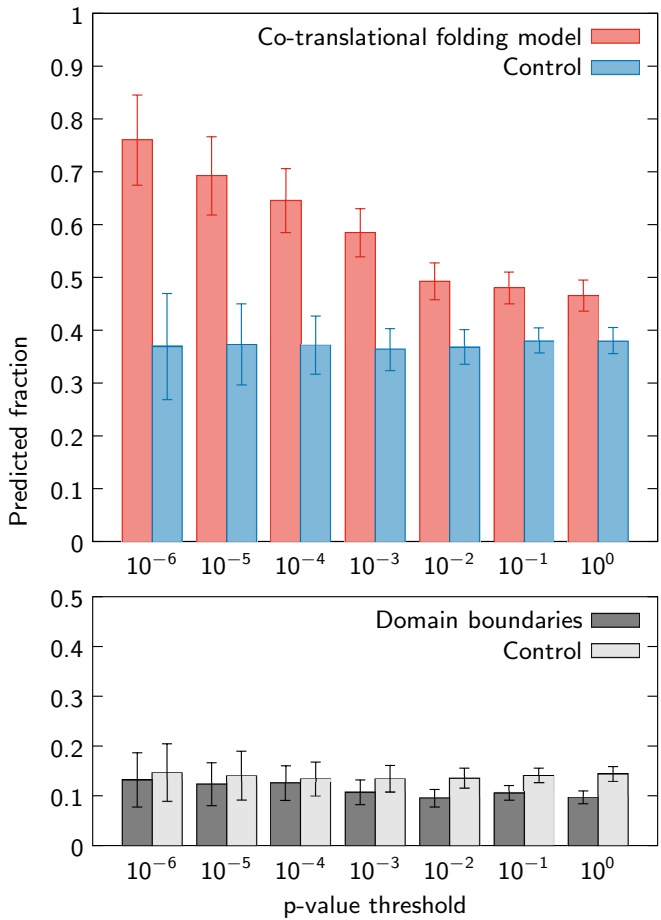}
  \caption{Above, the fraction of conserved, rare-codon enriched regions that follow a predicted co-translational folding intermediate increases as false positives are systematically eliminated.  In contrast, folding intermediates precede a consistently smaller fraction of the uniformly distributed enriched regions in randomized sequences.  Below, analyzing domain boundaries instead of folding intermediates similarly exhibits no dependence on the p-value threshold and accounts for a much lower percentage of the observed rare-codon enrichment loci.  The error bars on the control distributions indicate the standard deviation of 100 randomizations, while the error bars on the genomic data are estimated from binomial distributions at each p-value threshold.}
  \label{fig:explained}
\end{figure}

\subsection*{Conserved, enriched regions associate with predicted co-translational folding intermediates}

By applying this analysis to a set of approximately 500 \Ecoli{} proteins with known native structures, we find widespread support for our co-translational folding hypothesis.
In particular, we find that the co-translational folding intermediates predicted by our coarse-grained model account for a significant proportion ($\gtrsim 50\%$) of the putative slowly translated regions (\figref{fig:explained}).
Most importantly, we find that the fraction of rare-codon-enriched regions that can be explained by our model increases consistently as we reduce the p-value threshold for establishing evolutionary conservation.
In other words, the predictive power of our model improves as false positives related to the random clustering of rare codons are preferentially eliminated.
This trend is also robust with respect to variations in the precise definition of codon rarity (SI Appendix, Fig.~S7).

We further tested the sensitivity of our co-translational folding predictions by repeating the above analysis with randomized control sequences, which preserve the total number of pause sites at each p-value threshold but uniformly distribute their locations across the transcripts (\figref{fig:explained}; see Methods).
Although a significant fraction ($\sim\!35$\%) of the fictitious pause sites in the randomized sequences can also be explained by our model, likely due to chance overlaps with predicted intermediates, the difference between the genomic and randomized data increases markedly at lower p-value thresholds (one-sided ${p < 10^{-7}}$ at neutral-model p-value thresholds below 0.01; see SI Appendix, Fig.~S8a).
Two alternative controls (SI Appendix, Fig.~S9), in which the randomized pause sites are drawn from a non-uniform distribution with a 3'-end bias or obtained directly from reverse translations, verify that our results are not solely a consequence of the 3'-end rare-codon bias in the mRNA transcripts or the amino-acid sequences of the proteins.

\begin{figure}
  \includegraphics[width=\linewidth]{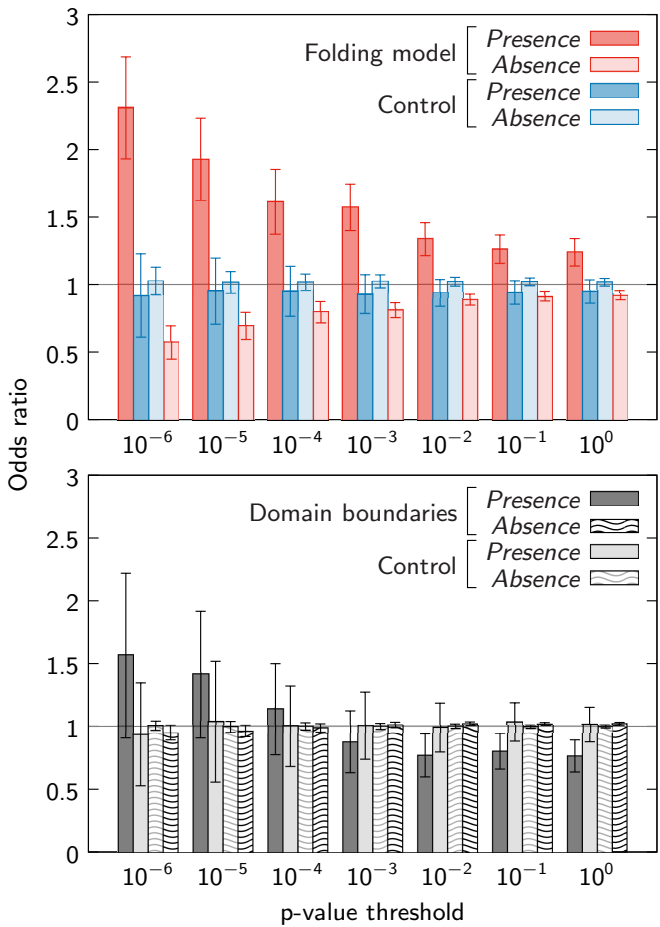}
  \caption{Conserved regions of rare-codon enrichment are more likely to appear between 20 and 60 codons downstream of a co-translational folding intermediate than elsewhere in a mRNA transcript.  Above, the odds ratio of finding an enriched region downstream of a predicted intermediate (presence) or downstream of no predicted intermediate (absence).  Unlike the comparisons with randomized control sequences, both ratios deviate significantly from unity and depend on the p-value threshold used.  Below, domain boundaries do not exhibit statistically significant associations with conserved pause sites.  Error bars are defined as in \figref{fig:explained}.}
  \label{fig:odds}
\end{figure}

Next, we performed inverse tests to assess whether co-translational folding intermediates are preferentially associated with putative translational pauses.
However, because the formation of an intermediate is not in itself a sufficient condition for a translational pause to be beneficial, we find that the overall frequency of such associations is small relative to the number of predicted intermediates (SI Appendix, Fig.~S10).
We therefore computed the odds ratio of finding conserved, rare-codon-enriched regions just downstream of a predicted intermediate, as opposed to elsewhere in a mRNA transcript.
The results shown in \figref{fig:odds} confirm that the association between folding intermediates and translational pause sites is highly significant (one-sided ${p < 10^{-7}}$ at neutral-model p-value thresholds below 0.01; see SI Appendix, Fig.~S8b) and, importantly, is not related to the overall frequency of predicted co-translational intermediates.
Here again, the predictive power of our model shows a strong dependence on the p-value threshold used for screening putative pause sites.
In contrast, tests with randomized control sequences do not deviate from an odds ratio of unity.

We also applied our analysis to structural domain boundaries, which have previously been suggested to play a role in coordinating co-translational folding~\cite{purvis1987efficiency}.
Nevertheless, in agreement with more recent works~\cite{jacobson2016quality}, we find little evidence of selection for translational pausing at domain boundaries.
For these comparisons, we used domain definitions for approximately 800 \Ecoli{} proteins from the SCOP database~\cite{murzin1995scop}.
\figref{fig:explained} shows that domain boundaries explain a much smaller fraction ($\lesssim 12\%$) of the putative pause sites than our folding model.
Furthermore, the predictive power of the domain-boundary hypothesis does not vary with the p-value threshold, and the odds ratios are nearly indistinguishable from the randomized controls (\figref{fig:odds}).
These conclusions also hold for various related hypotheses: instead of assuming that a domain must be completely synthesized before folding, we tested models where native structure begins to form either at a fixed number of residues prior to the domain boundary or at a fixed percentage of the domain length (SI Appendix, Fig.~S11).
In all cases, the correspondence between the domain boundaries and the conserved, rare-codon enriched loci is significantly weaker than the results of our co-translational folding model.
While these findings do not imply that domain boundaries are irrelevant for co-translational folding, we can conclude that the domain-boundary hypothesis is insufficient to explain the vast majority of conserved, slowly translated regions.

\section*{Discussion}

By integrating a multiple-sequence analysis of synonymous codon conservation with protein-folding theory, we have shown that highly conserved rare-codon clusters preferentially associate with predicted co-translational folding intermediates.
The putative pause sites in the \Ecoli{} genome that are both evolutionarily conserved and unaccounted for by the neutral model systematically appear downstream of predicted co-translational folding intermediates at distances that are similar to the length of the ribosome exit tunnel.
Our large-scale study therefore supports the hypothesis that beneficial pauses during protein synthesis follow key steps in the assembly of native structure.
Comparisons with randomized control sequences confirm that our observations are highly significant.

This analysis of co-translational folding pathways, as opposed to elements of the static native structure, provides new insights into the interplay between translation and the self-assembly kinetics of nascent proteins.
The stabilization of a partial structure often occurs well before a native domain is completely synthesized, especially in cases where the domain comprises more than 200 residues.
In particular, sub-domain co-translational folding intermediates typically appear when sufficient tertiary contacts are available to compensate for the loss of chain entropy that is required for folding.
Overall, a relatively small fraction ($\lesssim 15\%$) of all predicted intermediates are followed by conserved translational pauses (SI Appendix, Fig.~S10), but the association between folding intermediates and conserved pauses is highly significant (\figref{fig:odds}).
This observation is consistent with our hypothesis, since the effect of a pause depends on the relative timescales for translation and folding, as well as potential interference due to non-native interactions.
In addition, this observation explains why pause sites are not preferentially associated with domain boundaries: although fully synthesized domains may be stable on the ribosome, the prior formation of a partial-chain intermediate is likely to affect the subsequent folding rates for other parts of the protein.
As a result, the entire co-translational folding pathway must be considered when interpreting the effect of a pause in translation.
We anticipate that an optimal translation protocol~\cite{sharma2016physical} could be predicted with knowledge of the substructure-specific folding and translation rates, as well as their propensities for forming non-native interactions, including interactions with the surface of the ribosome~\cite{nilsson2017cotranslational}.
In addition, an optimal translation protocol is likely to be affected by the presence of misfolded intermediates, which may be avoided by increasing the local translation rate~\cite{obrien2014kinetic,trovato2017fast}.

The approach that we have taken in this work improves upon earlier studies of rare-codon usage, which have addressed alternative hypotheses regarding translational pausing but yielded mixed results~\cite{chaney2015roles,thanaraj1996ribosome,pechmann2013evolutionary,lee2010translationally,chartier2012large,brunak1996protein,zhou2009translationally}.
In addition to our distinct focus on co-translational folding pathways, our conclusions are more robust due to our use of a multiple-sequence analysis to detect evolutionary conservation, as well as our formulation of a neutral model that controls for both amino-acid composition and the inherent codon-usage variability across genes.
The statistical significance of our results is further increased by the much larger sample size used here.

While this paper was under review, we became aware of a contemporaneous study~\cite{chaney2017widespread} that identifies conserved rare-codon clusters via a complementary statistical analysis.
The authors also observe extensive rare-codon conservation across mRNA transcripts and similarly find no evidence of enrichment near domain boundaries.

Synonymous substitutions can also affect protein synthesis through mechanisms that are unrelated to protein folding, most notably via changes to mRNA secondary structure and stability~\cite{plotkin2011synonymous}.
However, many experimental studies have shown that these effects originate predominantly from substitutions near the 5'-end of the mRNA transcripts, and typically modulate the total protein production as opposed to the protein quality~\cite{kudla2009coding,goodman2013causes}.
Such mRNA-specific effects are thus a likely explanation for the observed 5'-end bias in rare-codon enrichment, where variations in translation speed are unlikely to play a role in co-translational folding.
Consequently, we have excluded N-terminal rare codons from our analysis.
In addition to rare-codon usage, various studies have proposed that additional factors, such as interactions between the nascent chain and the ribosome exit tunnel~\cite{chaney2015roles} or the presence of internal Shine--Dalgarno motifs~\cite{li2012anti}, can affect translation rates.
It is likely that a more complete picture of sequence-dependent translation kinetics will enable further refinements to the co-translational folding model presented here.

In conclusion, our study highlights the importance of optimal kinetic pathways for efficient biomolecular self-assembly.
Although a protein's amino-acid sequence entirely determines its thermodynamically stable structure, it is becoming increasingly clear that synonymous mutations are not always silent.
Our analysis provides strong evidence that evolutionary selection has tuned local translation rates to improve the efficiency of co-translational protein folding.
Further work is needed to understand the relationship between genome-wide codon usage and translation rates and to improve the prediction of co-translational folding intermediates, including those that contain significant amounts of non-native structure.
Nevertheless, our results indicate that folding kinetics play a role in evolutionary selection and suggest that similar relationships may exist for other biological self-assembly phenomena, such as the assembly of macromolecular complexes.

\section*{Methods}

We constructed alignments based on the amino-acid sequences of homologous genes from 18 prokaryotic species with between 50 and 85\% average amino-acid sequence identity to \Ecoli{} (SI Appendix, Table~S2).
We then computed p-values associated with rare-codon enriched regions, assuming biased reverse translations and a gene-specific model for the probability of each amino-acid type being encoded by a rare codon.
Consensus crystal structures were constructed for 511 non-membrane \Ecoli{} proteins with 500 residues or fewer using Protein Databank~\cite{berman2000protein} entries containing complete structures for sequences with at least 95\% amino-acid identity to the \Ecoli{} gene.
SCOP domain assignments were obtained from~\cite{murzin1995scop} for all proteins with at most 500 residues.
Due to the uncertainty in the number of amino acids that are concealed in the ribosome exit tunnel and the potential for steric interactions between folding intermediates and the ribosome, we consider a rare-codon-enriched region to be associated with a folding intermediate if the enriched region is anywhere between 20 and 60 codons downstream from the position at which an intermediate first becomes stable, ignoring enriched regions within the first 80 codons of a transcript.
An intermediate is identified whenever the monotonic co-translational free-energy profile decreases by more than ${1~\kB T}$ relative to the previous free-energy plateau; for example, see the pattern of alternating plateaus and precipitous free-energy decreases in the lower panel of \figref{fig:cotranslational_folding}.
To generate the randomized control sequences from which the control distributions in \figsref{fig:explained} and~\ref{fig:odds} were calculated, we sampled locations for fictitious rare-codon-enriched regions from a uniform distribution over each mRNA transcript, excluding the first 80 codons.
This uniform distribution was normalized such that the expected number of fictitious enriched regions is equal to the total number of observed enriched regions at each p-value threshold.
See SI Appendix, Sec.~S1 for complete details of all methodologies.
Essential data are provided in the SI Materials.
All code necessary to reproduce these results is available at \texttt{\footnotesize https://faculty.chemistry.harvard.edu/shakhnovich/software}.

\begin{acknowledgments}
  We thank Sanchari Bhattacharyya and Michael Manhart for many insightful discussions.
  This work was supported by NIH grants R01GM124044 and F32GM116231.
\end{acknowledgments}

\clearpage


\renewcommand{\thetable}{S\arabic{table}}
\renewcommand{\theequation}{S\arabic{equation}}
\renewcommand{\thesection}{S\arabic{section}}
\setcounter{table}{0}
\setcounter{equation}{0}
\setcounter{section}{0}
\setcounter{page}{1}

\onecolumngrid
\section*{Supplementary Information}
\vskip0.5ex
\twocolumngrid

\section{Extended methods}

\renewcommand{\thefigure}{\arabic{figure}}
\renewcommand{\figurename}{Supplementary Methods Figure}
\setcounter{figure}{0}

\subsection{Neutral model and statistics of rare-codon enrichment}
\label{sec:neutral_model}

In this section, we describe a neutral model of gene-specific synonymous codon usage.
For each genome, we define a set of rare codons that are used significantly less frequently than alternative synonymous codons in the most highly expressed genes~\cite{sharp1987codon}.
Using experimentally determined protein abundances to account for the differing expression levels among genes, we compute the relative synonymous codon frequencies for each amino-acid type,
\begin{equation}
  \label{eq:pusage}
  p^s_{\text{use}}(c | a) = \frac{\sum_g x_{sg}  \sum_{i = 1}^{L_{sg}} \bm{1}(c_{sgi} = c)}
  {\sum_g x_{sg} \sum_{i = 1}^{L_{sg}} \bm{1}(a_{sgi} = a)},
\end{equation}
where $c$ is a codon for an amino acid $a$, $x_{sg}$ is the protein abundance of gene $g$ in genome $s$, $i$ is an index that runs over all coding positions up to the protein length $L_{sg}$, and ${\bm{1}(\cdot)}$ is the indicator function.
We then define the set of rare codons as those codons whose protein-abundance-weighted relative usage, ${p^s_{\text{use}}(c|a)}$, is less than 10\%.
The relative usages of \Ecoli{} rare codons, as determined by \eqref{eq:pusage}, are shown in \tabref{tab:rare_codons}.
We use a composite data set for the protein abundances in \Ecoli{}~\cite{wang2012paxdb} and assume that these abundances are similar for all prokaryotic genomes in our alignment (\tabref{tab:genome_list}).
The rare-codon definitions turn out to be comparable, but not identical, for these genomes.

Next, we address the absolute enrichment of rare codons in a local region of a gene, without considering the amino-acid sequence.
We calculate the average fraction of rare codons in the $M$ aligned sequences, which all have at least 50\% amino-acid identity with respect to the \Ecoli{} gene and differ in length by no more than 20\%,
\begin{equation}
  \lambda_g \equiv M^{-1} \sum_{s=1}^M L_{sg}^{-1} \sum_{i = 31}^{L_{sg}} \bm{1}[p^{sgi}_{\text{use}}(c | a) \le 0.1].
\end{equation}
Here and below, we exclude the first 30 codons to avoid the 5'-end bias, and we define ${p^{sgi}_{\text{use}}(c | a) \equiv p^s_{\text{use}}(c_{sgi} | a_{sgi})}$ for brevity.
We then define a local-region enrichment threshold $n^{sg}_{\text{enr}}$ for each aligned sequence by considering a Poisson process with rare-codon-usage rate~$\lambda_g$,
\begin{equation}
  n^{sg}_{\text{enr}} \equiv \min\left( l > 0 \, | \, 1 - p^{\phantom\alpha}_{\text{Poisson}}(\lambda_g, l, n_{s}) \le 0.15 \right),
  \label{eq:nenr}
\end{equation}
where ${p^{\phantom\alpha}_{\text{Poisson}}(\lambda, l, n)}$ is the cumulative distribution function of $l$ events occurring in $n$ trials given a rate $\lambda$.
The cumulative probability $0.15$ corresponds to an approximately one-standard-deviation fluctuation in the local rare-codon usage.
Every local region has a fixed width within the \Ecoli{} sequence, in which case ${n_s = n}$.
(We have chosen ${n = 15}$ in the main text.)
However, due to insertions and deletions, the number of codons that align to a given $n$-codon region of the \Ecoli{} sequence may be different for the other sequences in the alignment.
Therefore, when applying \eqref{eq:nenr} to these other sequences, we use the number of codons $n_s$ that align to each $n$-codon region of the \Ecoli{} sequence, including any insertions.
With this approach, we can identify structurally equivalent regions that are enriched in rare codons in each aligned sequence.

We can now calculate a p-value that accounts for biases in rare-codon usage due to the local amino-acid composition.
To do so, we estimate the probability of using a rare codon for each amino-acid type in a given gene by analyzing the relative codon-usage bias of all genes with a similar overall rare-codon usage~$\lambda$,
\begin{equation}
  \label{eq:prare}
  p^{s,\lambda}_{\text{rare}}(a) \equiv \!
  \frac{\sum_{\{g\}_\lambda} \!\sum_{i = 31}^{L_{sg}} \bm{1}(a_{sgi} = a) \bm{1}[p^{sgi}_{\text{use}}(c | a) \le 0.1]}
       {\sum_{\{g\}_\lambda} \!\sum_{i = 31}^{L_{sg}} \bm{1}(a_{sgi} = a)},
\end{equation}
where ${\{g\}_\lambda}$ is the set of genes with similar~$\lambda$.
In practice, we implement \eqref{eq:prare} by sorting all genes according to their overall rare-codon usages and splitting them into ten groups, each comprising approximately 400 genes.
However, because the average codon usage is not exactly the same for all genes in each of these groups, we then adjust the rare-codon probabilities slightly to match $\lambda_g$,
\begin{equation}
  p^{sg}_{\text{rare}}(a) = \gamma p^{s,\lambda_g}_{\text{rare}}\!(a),
\end{equation}
where the scaling factor $\gamma$ is chosen such that
\begin{equation}
  \label{eq:gamma}
  M^{-1} \! \sum_{s=1}^M L_{sg}^{-1} \sum_{i = 31}^{L_{sg}} \!
  \frac{\gamma p^{s,\lambda_g}_{\text{rare}}\!(a_{sgi})}
       {1 + (\gamma - 1) p^{s,\lambda_g}_{\text{rare}}\!(a_{sgi})}
       = \lambda_g.
\end{equation}
\eqref{eq:gamma} is nearly linear in $\gamma$ while ensuring that all probabilities are positive.

Finally, the neutral-model probability that a local region contains at least $n^{sg}_{\text{enr}}$ rare out of $n_{s}$ codons is
\begin{equation}
  p^{sg}_{\text{enr}} = \!\! \sum_{l = n^{sg}_{\text{enr}}}^{n_{s}} \!\! \sum_{\Omega}^{\text{perm.}} \!
  \prod_{i = 1}^{l} p^{sg}_{\text{rare}}(a_{s\Omega_i}) \!\!
  \prod_{j = l + 1}^{n_{s}} \! [1 - p^{sg}_{\text{rare}}(a_{s\Omega_j})],
  \label{eq:penr}
\end{equation}
where ${\{\Omega\}}$ is the set of all permutations of the codon positions within the given region and $\Omega_i$ is the position at index $i$ in the permutation $\Omega$.
The probability of observing at least $m_{\text{obs}}$ enriched sequences is then
\begin{equation}
  \label{eq:pneutral}
  p^g_{\text{neu}}(m_{\text{obs}},M) = \!\!\! \sum_{m = m_{\text{obs}}}^M \!\!\! \sum_{\sigma}^{\text{perm.}} \!
  \prod_{i = 1}^{m} p^{\sigma_i g}_{\text{enr}} \!\!\! \prod_{j = m + 1}^{M} \!\! (1 - p^{\sigma_j g}_{\text{enr}}),
\end{equation}
where ${\{\sigma\}}$ is the set of all orderings of the $M$ sequences, and $\sigma_i$ is the sequence at the $i$th index of the ordering~$\sigma$.
We require ${m_{\text{obs}} / M \ge 0.75}$ for an enriched region to be considered conserved (see \figref{fig:alternative_conservation}).
\eqref{eq:pneutral} is the neutral-model p-value used throughout this work.

\subsection{Quasi-equilibrium co-translational folding model}
\label{sec:folding_model}

\begin{figure}
  \includegraphics{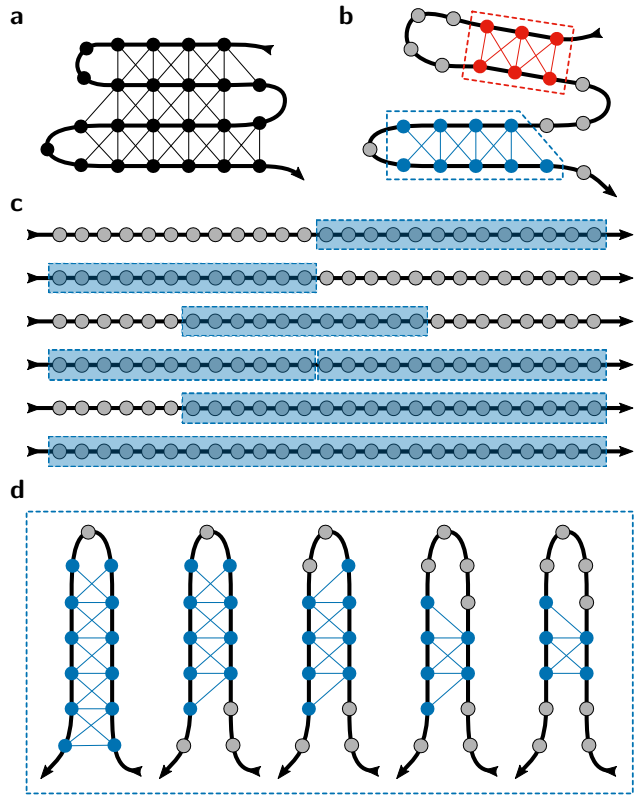}
  \caption{Schematic of the coarse-grained model (adapted from \refcite{jacobs2016structure}).  (a) Native contacts (thin black lines) are drawn between residues (circles) that are connected by the polymer backbone (thick black line).  (b) A partially folded configuration may contain one or more structured regions.  In this example configuration, native contacts are formed within two such regions (red and blue sets of residues), while the disordered gray residues do not form any native contacts; each structured region is associated with a single loop of either four disordered residues (red region) or one disordered residue (blue region).  (c) Example sets of non-overlapping structured regions (dashed blue boxes).  For each structured region, we minimize the mean-field free energy (\eqref{eq:Fmf}); the total free energy is the sum of the minimized mean-field free energies of all structured regions.  (d)~An example ensemble of similar configurations.  The configurations differ by the number of residues in the structured region as well as the length, but not the position, of the single disordered loop.}
  \label{fig:folding_model}
\end{figure}

In this section, we adapt the native-centric coarse-grained model described in \refcite{jacobs2016structure} to predict free-energy landscapes for co-translational folding (\smfigref{fig:folding_model}a).
A native contact between a pair of residues is formed when at least one pair of heavy atoms is separated by at most 4~{\AA} in the consensus crystal structure and the residues are at least three positions apart in the primary sequence.
The effective potential energy between residues $u$ and $v$ depends on the number of such heavy-atom contacts, ${n^{\text{nc}}_{uv} \ge 1}$; whether the pair of residues forms a hydrogen bond, $\bm{1}^{\text{hb}}_{uv}$; and whether the contact is part of an $\alpha$-helix, $\bm{1}^{\text{helix}}_{uv}$:
\begin{equation}
  \label{eq:epsilon}
  \epsilon_{uv} = -\left(\alpha_{\text{helix}}\right)^{\bm{1}_{uv}^{\text{helix}}} \! \left[n_{uv}^{\text{nc}}
    + \alpha_{\text{hb}}\bm{1}_{uv}^{\text{hb}}\right],
\end{equation}
with ${\alpha_{\text{helix}}=5/8}$ and ${\alpha_{\text{hb}}=16}$.

In this discrete model, a microscopic configuration is defined by the set of native contacts that are formed.
The free energy of such a configuration is determined by opposing energetic and entropic contributions: native contacts are stabilizing according to \eqref{eq:epsilon}, and configurational entropy is gained when native contacts are broken.
Furthermore, sets of native contacts can be decomposed into independent structured regions (\smfigref{fig:folding_model}b).
Within each structured region, all possible native contacts are formed among the participating residues; however, different structured regions are linked by chains of disordered residues and can thus move independently.
We therefore evaluate the free energy of each structured region separately.
The free energy of a configuration $g$ with a single structured region is
\begin{equation}
  \label{eq:Fg}
  \frac{f(g)}{\kB T} =
  \frac{1}{2} \sum_{u,v \in g} \frac{\epsilon_{uv}}{\kB T}
  + \frac{\mu}{\kB T} (N_g - 1) - \frac{\Delta S_{\text{l}}(g)}{\kB},
\end{equation}
where $u$ and $v$ are residues that form native contacts in the configuration~$g$, $N_g$ is the number of contact-forming residues, $\kB$ is the Boltzmann constant, $T$ is the absolute temperature, and ${\mu = 2 \kB T}$ is the entropic penalty associated with native-contact formation~\cite{jacobs2016structure}.
Sequences of disordered residues that begin and end in the same structured region are referred to as loops (\smfigref{fig:folding_model}b--c).
The total loop entropy, ${\Delta S_{\text{l}}(g)}$, is the sum of ${\Delta S_{\text{l}}(l)}$ for all loops~${\{l\}_g}$ that terminate in the structured region,
\begin{equation}
  \label{eq:Sloop}
  \frac{\Delta S_{\text{l}}(l)}{\kB} \equiv
  \begin{cases}
    -\frac{|l| \mu}{\kB T}
    & \!\!\!\text{if } |l| \le b, \\
    -\frac{b \mu}{\kB T} - \frac{3}{2} \! \left[ \ln \frac{|l|}{b} + \frac{r(l)^2}{b^2 |l|} \right]
    & \!\!\!\text{if } |l| > b, \quad
  \end{cases}
\end{equation}
where ${|l|}$ is the number of disordered residues in the loop, ${r(l)}$ is the distance between the loop endpoints, and we assume Gaussian statistics for all loops longer than the Kuhn length ${b = 2}$ residues.

Our goal is to compute the free energy of a nascent chain of length $L$, ${F(L)}$.
First, we note that the free energy of a polypeptide chain in this model is equal to the sum of the free energies of one or more structured regions, which do not interact with one another.
To minimize the free energy ${F(L)}$, any combination of non-overlapping structured regions using the first $L$ residues is permitted.
For example, various possible sets of structured regions are illustrated in \smfigref{fig:folding_model}c: native contacts are formed within each of the structured regions in a configuration, but no contacts are formed between residues in different structured regions.
The set of allowed sets of structured regions can be written as
${\bm{r} \!\equiv \!\{(a_1,b_1), \ldots, \!(a_k,b_k) |\, a_1 \!\ge\! 0, a_i \!\ge\! b_{i - 1} \forall i \!\in\! [2,k], b_k \!\le\! L\}}$,
where the pair ${(a_i,b_i)}$ indicates the first and last residues that can participate in the structured region with index~$i$, and the number of structured regions is ${k \ge 0}$.
Then, to account for the fact that many distinct microstates are typically consistent with a given structured region, we write the free energy of a nascent chain with a specific set of structured regions $\bm{r}$ as
\begin{eqnarray}
  \frac{F_{\bm{r}}(L)}{\kB T} &=& -\!\! \sum_{(a,b)\in\bm{r}} \frac{F_{ab}}{\kB T} \\
  &=& -\!\! \sum_{(a,b)\in\bm{r}} \!\! \ln \left[ \sum_{\{g\}_{ab}} e^{-F(g) / \kB T} \right], \nonumber
  \label{eq:FLr_SI}
\end{eqnarray}
where ${\{g\}_{ab}}$ indicates the set of all microstates that have a single structured region incorporating residues in the range ${[a, b]}$, and ${F_{ab}}$ is the free energy of this ensemble.
Unfortunately, directly calculating ${F_{ab}}$ requires a Monte Carlo approach~\cite{jacobs2016structure}, which can be computationally intensive.

We therefore introduce a mean-field approximation to compute the free energy of a single structured region, ${F_{ab}}$.
We assume that all configurations in this ensemble contain a single structured region with an essentially equivalent set of loops (i.e., the loops grow or shrink by adding or removing adjacent residues, but otherwise the structured region remains unchanged; see \smfigref{fig:folding_model}d).
Instead of considering the set of discrete configurations ${\{g\}}$, we now characterize the ensemble by the average ordering of each residue in this set of microscopic configurations, ${\{\rho_u\}}$.
Ignoring correlations in the residues beyond nearest-neighbor native contacts, we can write the mean-field free energy of this ensemble of microscopic configurations as the sum of the average energetic and entropic contributions,
\begin{eqnarray}
  \frac{F_{\text{mf}}(\{\rho_u\})}{\kB T} &=& \frac{U_{\text{mf}}}{\kB T} - \frac{S_{\text{mf}}}{\kB} \\
  \label{eq:Fmf}
  &=& -\frac{1}{2} \sum_{u,v} \frac{\rho_u \tilde\epsilon_{uv} \rho_v}{\kB T} - \sum_{u} \ln q_u \\
  &\,& \qquad - \frac{\langle \Delta S_{\text{l}} \rangle_{\{\rho_u\}}}{\kB} \nonumber,
\end{eqnarray}
where $q_u$ is the single-residue mean-field partition function
\begin{equation}
q_u = 1 + \exp\left(-\tilde\mu - \sum_{v \in g} \tilde\epsilon_{uv} \rho_v \right).
\end{equation}

Native contacts are correlated on a length scale comparable to the Kuhn length due to the finite size of the residues and the chain connectivity~\cite{jacobs2016structure}.
We have excluded native contacts between residues that are fewer than three positions apart in the sequence when constructing the energy matrix~${\{\epsilon_{uv}\}}$, because such contacts are likely to be present in both the denatured and folded ensembles.
Nevertheless, we need to account for local geometric correlations in native contact formation.
To impose correlations between residues that are very close in the sequence, we add nearest and next-nearest-neighbor couplings to ${\{\epsilon_{uv}\}}$ and ${\{\mu_u\}}$,
\begin{eqnarray}
  \tilde\epsilon_{uv} &=& \epsilon_{uv}
  - \bm{1}_r(v \!+\! 1) \!\left[\delta_{u,v+1} \lambda_1 - \bm{1}_r(v \!+\! 2) \delta_{u,v+2} \lambda_2 \right]\!, \;\qquad\\
  \tilde\mu_{u} &=& \mu + \bm{1}_r(v \!+\! 1) \left[\lambda_1 + \bm{1}_r(v \!+\! 2) \lambda_2\right],
\end{eqnarray}
where $\delta$ is the Kronecker delta, ${\bm{1}_r(v)}$ indicates whether the residue $v$ can form native contacts in the structured region, and we assume that ${u < v}$ and ${\bm{1}_r(u) = 1}$ for notational simplicity.
We have chosen coupling constants ${\lambda_1 = 0.8}$ and ${\lambda_2 = 0.4}$ to obtain comparable results to the Monte Carlo simulations reported in \refcite{jacobs2016structure}.
Choosing ${\lambda_1, \lambda_2 > 0}$ tends to increase the average ordering at the free-energy minimum for all residues that form multiple contacts in the structured region.

\begin{figure}[ht!]
  \includegraphics{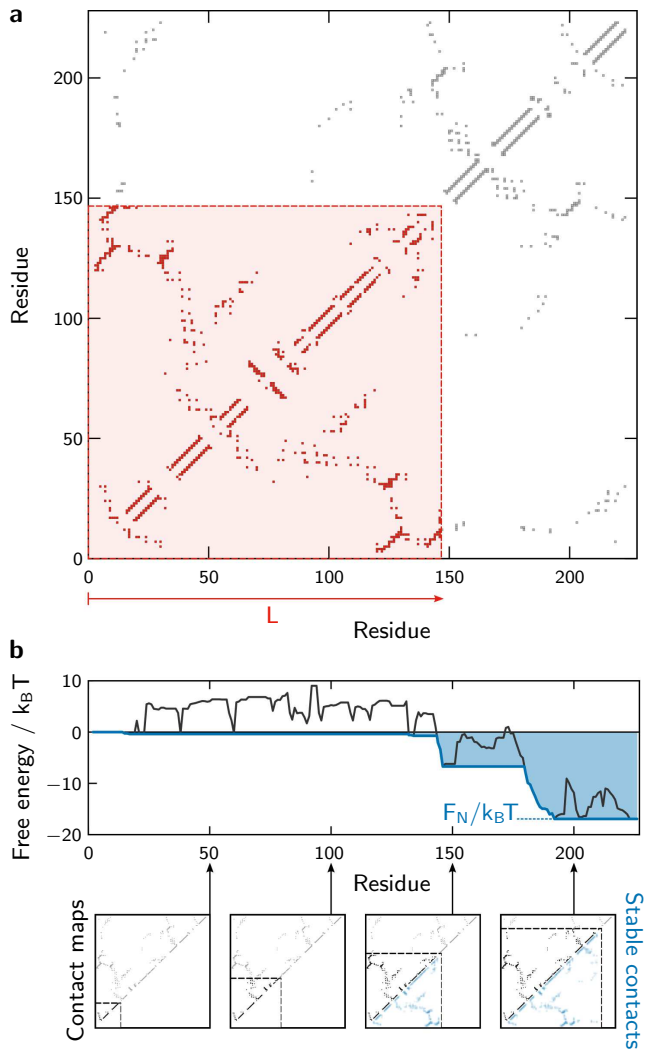}
  \caption{Calculation of the nascent-chain free energy from the native-state contact map.  (a) An example consensus contact map for the \Ecoli{} gene \textit{cmk}.  The highlighted region shows the native contacts that can be formed by a nascent chain of length $L$.  (b) The corresponding free-energy calculation, reproduced from Figure~2, shows the minimum free energy, ${F(L)}$, for each nascent-chain length (blue curve; \eqref{eq:FL_SI}).  To better illustrate the contributions to this calculation, we also plot the minimum mean-field free energy obtained when the $L$th residue is included in one of the structured regions (black curve), i.e., ${b_i = L}$ for one of the structured regions ${(a_i, b_i)}$ in the set of structured regions~${\bm{r}}$.  The free energy of this artificially constrained ensemble is a non-monotonic function of the chain length, since newly synthesized residues may not be able to join the stable tertiary structure until additional, downstream residues become available.  However, ${F(L)}$ can be determined by finding the most negative point on this black curve at any chain length less than or equal to $L$.  The free energy of the full-length chain is $F_N$.  Contact maps are shown for select nascent-chain lengths, where the stable contacts are indicated below the diagonal.}
  \label{fig:contact_map}
\end{figure}

To determine ${F_{\text{mf}}}$ and ${\{\rho_u\}}$, we first minimize the mean-field free energy without considering the loop term, ${\Delta S_{\text{l}}}$.
\eqref{eq:Fmf} can be minimized numerically using a standard non-linear, multi-dimensional minimization algorithm.
The probability of forming a native contact between residues $u$ and $v$ in the mean-field model is then ${p^{\text{contact}}_{uv} = \rho_u \bm{1}(\epsilon_{uv} < 0) \rho_v}$; example contact maps showing ${\{p^{\text{contact}}_{uv}\}}$ at the free-energy minimum can be found in \smfigref{fig:contact_map}.
While it is not guaranteed that \eqref{eq:Fmf} will have a single minimum, in practice we find that, excluding the trivial solution, ${\{\rho_u\} = 0}$, this is almost always the case.
Such behavior is expected from our previous study~\cite{jacobs2016structure} of the model presented in \eqref{eq:Fg}.

Next, we compute the ensemble-averaged loop entropy, ${\langle \Delta S_{\text{l}} \rangle_{\{\rho_u\}}}$, perturbatively by assuming that this contribution to \eqref{eq:Fmf} does not alter ${\{\rho_u\}}$ at the free-energy minimum.
To allow disordered loops to grow or shrink by adding or removing adjacent residues, we calculate the Boltzmann-weighted average loop entropy for all possible pairs of terminal residues ${\{(u,v)\}_l}$ for each loop $l$,
\begin{equation}
  \langle \Delta S_{\text{l}}(l) \rangle_{\{\rho_u\}} =
  \frac{\sum_{\{(u,v)\}_l} \rho_{u} p_{\text{l}}([u,\!v]) \rho_{v}
    e^{\frac{\Delta S_{\text{l}}\left([u,\!v]\right)}{\kB}} S_{\text{l}}\left([u,\!v]\right)}
       {\sum_{\{(u,v)\}_l} \rho_{u} p_{\text{l}}([u,\!v]) \rho_{v}
         e^{\frac{\Delta S_{\text{l}}\left([u,\!v]\right)}{\kB}}},
\end{equation}
where ${p_{\text{l}}([u,\!v]) \equiv \prod_{u > t > v}(1 - \rho_t)}$, and ${\Delta S_{\text{l}}\left([u,\!v]\right)}$ implies \eqref{eq:Sloop} computed with the loop endpoints ${(u,v)}$.

Finally, we can calculate the minimum free energy of a nascent chain in this mean-field approximation.
We consider native contacts among the first $L$ residues (\smfigref{fig:contact_map}a).
We then minimize the mean-field free energy by finding the lowest-free-energy set of structured regions,
\begin{equation}
  \frac{F(L)}{\kB T} \simeq \min_{\bm{r}} \!\left[ \sum_{(a,b)\in\bm{r}} \!\! \frac{F_{\text{mf}}(\{\rho_u\}_{ab})}{\kB T} \right],
  \label{eq:FL_SI}
\end{equation}
where ${\{\rho_u\}_{ab}}$ implies that ${\rho_u = 0}$ for all ${u < a}$ and ${u > b}$.
At each chain length, ${F(L)}$ reports the minimum free energy relative to a random coil of length $L$ (\smfigref{fig:contact_map}b).
The free energy defined in \eqref{eq:FL_SI} is used for identifying co-translational folding intermediates as described in the main text.

\subsection{Analysis of co-translational folding pathways}

Due to the uncertainty in the number of amino acids that are concealed in the ribosome exit tunnel and the potential for steric interactions between folding intermediates and the ribosome, we consider a rare-codon-enriched region to be associated with a folding intermediate if the enriched region is anywhere between 20 and 60 codons downstream from the position at which an intermediate first becomes stable.
We ignore enriched regions within the first 80 codons of a transcript, since the conservation of rare codons near the 5'-end of the mRNA transcripts is believed to originate from other factors, such as efficient translation initiation.
A co-translational folding intermediate is identified whenever the monotonic co-translational free-energy profile decreases by more than ${1~\kB T}$ relative to the previous free-energy plateau; for example, see the pattern of alternating plateaus and precipitous free-energy decreases in the lower panel of Figure~2.
In Figure~3, the `predicted fraction' is the number of rare-codon-enriched regions that are preceded by a folding intermediate, $n_{\text{RI}}$, divided by the total number of enriched regions, $n_{\text{R}}$.
To compute the odds ratios in Figure~4, we first determined the fraction, $f_{\text{I}}$, of all codons in our data set that are 20 to 60 positions downstream of a predicted intermediate.
The `presence' odds ratio is then defined as ${(n_{\text{RI}} / n_{\text{R}}) / f_{\text{I}}}$, and the `absence' odds ratio is ${(1 - n_{\text{RI}} / n_{\text{R}}) / (1 - f_{\text{I}})}$.

To generate the randomized control sequences from which the control distributions in Figures~3 and 4 were calculated, we sampled locations for fictitious rare-codon-enriched regions from a uniform distribution over each mRNA transcript, excluding the first 80 codons.
This uniform distribution was normalized such that the expected number of fictitious enriched regions is equal to the total number of observed enriched regions at each p-value threshold.

\clearpage

\onecolumngrid
\section{Supplementary tables and figures}

\renewcommand{\figurename}{FIG.}
\renewcommand{\thefigure}{S\arabic{figure}}
\setcounter{figure}{0}

\begin{table}[h]
  {
    \setlength{\tabcolsep}{2ex}
    \begin{tabular}{cccc}
      Amino acid & Codon & Rel.~usage & Max.~rel.~usage \\
      \hline \hline
      G & GGA & 0.0387 & 0.4474 (GGC) \\
      ~ & GGG & 0.0811 & ~ \\
      \hline
      I & ATA & 0.0161 & 0.6062 (ATC) \\
      \hline
      L & CTA & 0.0269 & 0.6785 (CTG) \\
      ~ & CTC & 0.0880 & ~ \\
      ~ & CTT & 0.0853 & ~ \\
      ~ & TTA & 0.0495 & ~ \\
      ~ & TTG & 0.0718 & ~ \\
      \hline
      P & CCC & 0.0842 & 0.6344 (CCG) \\
      \hline
      R & AGA & 0.0097 & 0.5283 (CGT) \\
      ~ & AGG & 0.0043 & ~ \\
      ~ & CGA & 0.0170 & ~ \\
      ~ & CGG & 0.0260 & ~ \\
      \hline
      S & AGT & 0.0539 & 0.3198 (TCT) \\
      ~ & TCA & 0.0547 & ~ \\
      ~ & TCG & 0.0657 & ~ \\
      \hline
      T & ACA & 0.0840 & 0.5224 (ACC)
    \end{tabular}
  }
  \caption{Rare codons in the \Ecoli{} genome.  A codon is considered to be rare if the abundance-weighted relative usage, as defined in \eqref{eq:pusage}, is less than 10\%.  For comparison, the highest relative-usage codon for each amino acid is shown in the rightmost column.}
  \label{tab:rare_codons}
\end{table}

\begin{figure}[h]
  \includegraphics{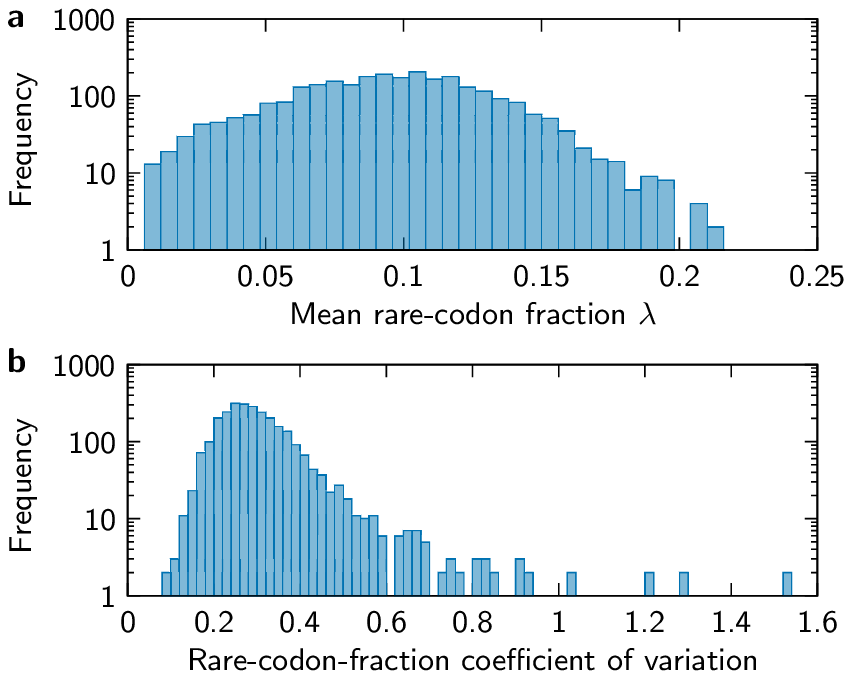}
  \caption{The overall rare-codon usage in the \Ecoli{} genome.  (a) \Ecoli{} genes exhibit a relatively broad range of average rare-codon usages,~$\lambda$.  (b)  However, for each gene, the rare-codon usage is remarkably similar across prokaryotic genomes.  The similarity among genomes is quantified here by the coefficient of variation (CV), defined as the ratio of the standard deviation to the mean rare-codon fraction across genomes.  The high-CV outliers are overwhelmingly associated with the lowest $\lambda$ genes.}
  \label{fig:homologous_adaptation}
\end{figure}

\begin{figure*}[h]
  \includegraphics{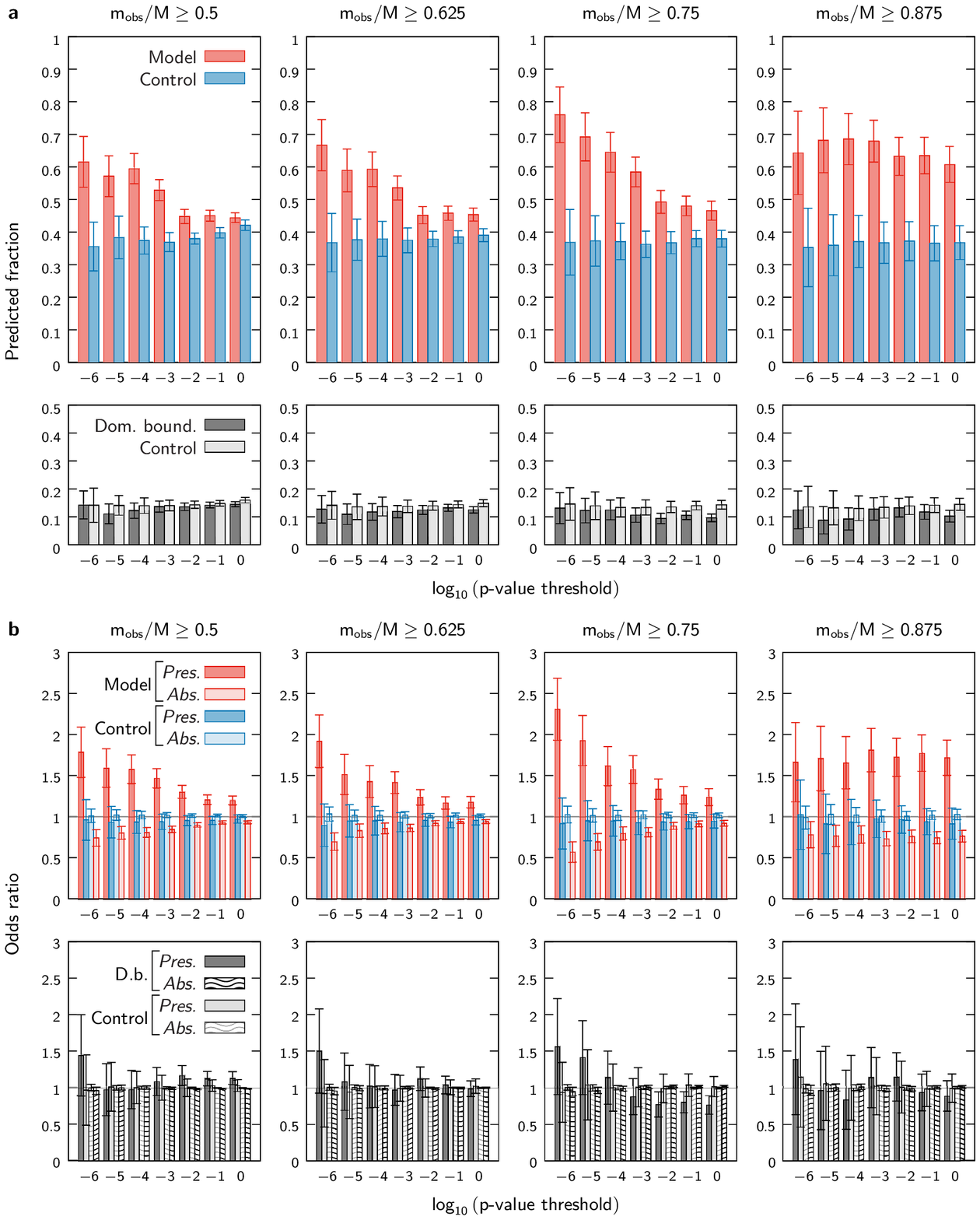}
  \caption{The accuracy of the co-translational folding model predictions does not depend strongly on the conservation cutoff used in determining putative pause sites.  In each column, we require that the fraction of the aligned sequences exhibiting a rare-codon enriched region, ${m_{\text{obs}} / M}$, equal or exceed a prescribed cutoff (see \secref{sec:neutral_model}); throughout this work, we have used a cutoff of 75\%.  Panels in (a) are analogous to Figure~3, while panels in (b) are analogous to Figure~4.  For the 87.5\% cutoff, almost all conserved regions are associated with very low p-values, and thus the folding model outperforms the control at all p-value thresholds; however, the scarcity of enriched regions at this strict conservation cutoff tends to reduce the statistical significance.}
  \label{fig:alternative_conservation}
\end{figure*}

\begin{table}[h]
  {
    \setlength{\tabcolsep}{2ex}
    \begin{tabular}{r|cc|cc|l}
      ~ & \multicolumn{2}{|c|}{Codon identity} & \multicolumn{2}{|c|}{Amino-acid identity} & ~ \\
      Genome ID & \Ecoli{} & Average & ~\Ecoli{}~ & Average & Species \\
      \hline \hline
      NC\_000913.3 & 1.000 & 0.390 & 1.000 & 0.682 & \textit{Escherichia coli} K-12 MG1655 \\
      NC\_007606.1 & 0.775 & 0.376 & 0.830 & 0.654 & \textit{Shigella dysenteriae} \\
      NZ\_CP007557.1 & 0.473 & 0.382 & 0.803 & 0.687 & \textit{Citrobacter freundii} \\
      NC\_003198.1 & 0.448 & 0.397 & 0.767 & 0.681 & \textit{Salmonella enterica} \\
      NC\_014121.1 & 0.420 & 0.380 & 0.739 & 0.675 & \textit{Enterobacter cloacae} \\
      NC\_015663.1 & 0.407 & 0.384 & 0.726 & 0.675 & \textit{Enterobacter aerogenes} \\
      NZ\_CP007215.1 & 0.406 & 0.375 & 0.722 & 0.673 & \textit{Enterobacter sacchari} \\
      NC\_016845.1 & 0.405 & 0.384 & 0.724 & 0.671 & \textit{Klebsiella pneumoniae} \\
      NC\_009778.1 & 0.384 & 0.372 & 0.694 & 0.659 & \textit{Cronobacter sakazakii} \\
      NZ\_CP009450.1 & 0.382 & 0.369 & 0.702 & 0.658 & \textit{Pluralibacter gergoviae} \\
      NZ\_CP009451.1 & 0.373 & 0.359 & 0.692 & 0.655 & \textit{Cedecea neteri} \\
      NC\_017910.1 & 0.349 & 0.342 & 0.658 & 0.631 & \textit{Shimwellia blattae} \\
      NZ\_CP009454.1 & 0.306 & 0.318 & 0.591 & 0.598 & \textit{Pantoea rwandensis} \\
      NC\_016818.1 & 0.304 & 0.310 & 0.590 & 0.592 & \textit{Rahnella aquatilis} \\
      NC\_014500.1 & 0.301 & 0.313 & 0.582 & 0.586 & \textit{Dickeya dadantii} \\
      NC\_008800.1 & 0.284 & 0.292 & 0.594 & 0.604 & \textit{Yersinia enterocolitica} \\
      NC\_003143.1 & 0.273 & 0.283 & 0.578 & 0.592 & \textit{Yersinia pestis} \\
      NC\_022546.1 & 0.268 & 0.289 & 0.502 & 0.520 & \textit{Plautia stali} symbiont (\textit{Enterobacter})
    \end{tabular}
  }
  \caption{The prokaryotic genomes used in the multiple-sequence alignments.  The middle columns show the average codon and amino-acid identities of all genes relative to both the \Ecoli{} genome and the other species in the alignment.  When computing codon and amino-acid identities, we have excluded genes whose lengths differ from the \Ecoli{} homolog by more than 20\%, and we have ignored all insertions and deletions.}
  \label{tab:genome_list}
\end{table}

\begin{figure}[h]
  \includegraphics{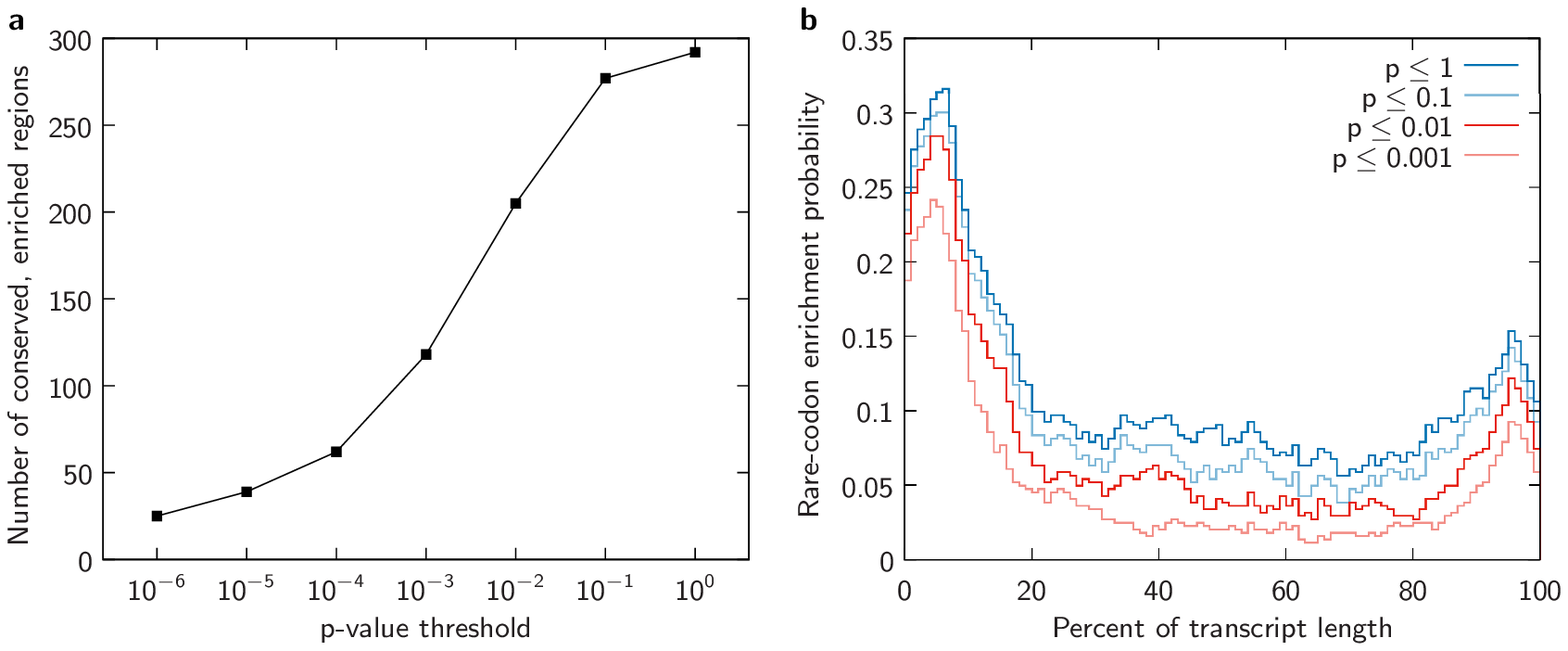}
  \caption{The distribution of conserved, rare-codon enriched regions in the 511-protein data set used in this study.  As discussed in the main text, we include only those regions that are conserved across at least 75\% of the aligned sequences.  (a) Excluding enriched regions within 80 codons of the N-terminus, the number of conserved, enriched regions at each p-value threshold (\eqref{eq:pneutral}).  (b) The distribution of enriched regions within mRNA transcripts at various p-value thresholds.  The locations within the transcripts are normalized by the lengths of the transcripts.  The apparent biases in the locations of the enriched regions towards both the 5' and 3'-ends of the transcripts become more significant as the p-value threshold is lowered.  In contrast, random reverse translations of the \Ecoli{} proteome using the gene-specific rare-codon probabilities (\eqsref{eq:prare}--\ref{eq:gamma}) yield uniform distributions at all p-value thresholds.  The bias towards rare-codon enrichment at the 3'-end is therefore not a consequence of the amino-acid composition near the C-terminus.}
  \label{fig:rc_distribution}
\end{figure}

\begin{figure}[h]
  \includegraphics{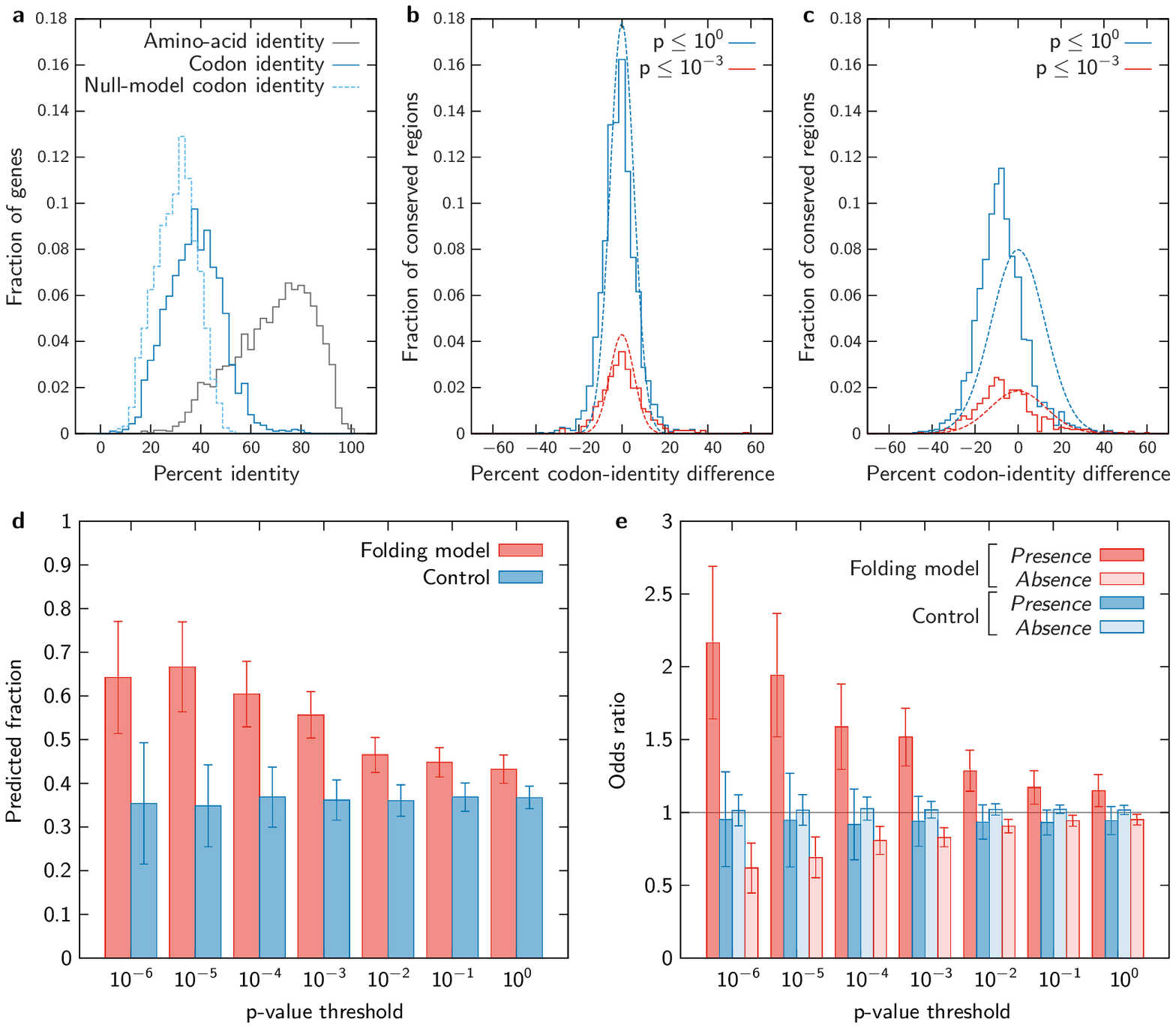}
  \caption{Analysis of the sequence divergence among the prokaryotic genomes listed in \tabref{tab:genome_list}.  (a)~Mutual-identity distributions for the genes considered in Figures~3 and~4.  For the purpose of this figure, we define the mutual codon identity of a multiple-sequence alignment to be ${\sum_{i=1}^L \sum_{r \ne s} \bm{1}(c_{si} = c_{ri}) / [L M (M - 1) / 2]}$, where $L$ is the number of positions aligned to the \Ecoli{} gene, $M$ is the number of sequences in the alignment, and $r$ and $s$ are the genome indices.  The mutual amino-acid identity is defined analogously.  The null-model codon-identity distribution (dashed line) is calculated from reverse translations of the aligned amino-acid sequences using the gene-specific synonymous codon probabilities described in \secref{sec:neutral_model}.  The observed codon-identity distribution is close to the null-model distribution, indicating that the synonymous substitutions are, in general, sufficiently diverse to perform a conservation analysis.  (b)~The distribution of the differences in mutual codon identity between each putative conserved region and the corresponding complete gene.  Results are shown for conserved regions at two conservation thresholds.  For comparison, we also show the expected distributions (dotted lines) that result from a uniform sampling of $n$-codon loci from each complete gene.  The similarity between the observed and expected distributions indicates that the overall synonymous codon divergence in the putative conserved loci is not significantly different from the remainder of each gene.  (c)~Examining only those specific positions in the putative conserved regions that have rare codons in the \Ecoli{} sequence does not affect this conclusion.  In fact, the average codon identity at these positions is typically lower than that of the gene as a whole, since these specific positions in the alignment are more likely to contain amino acids that are encoded by many synonymous codons.  Taken together, panels \textbf{a}--\textbf{c} indicate that, overall, our analysis identifies loci with a conserved enrichment of rare codons, as opposed to loci with reduced codon-level sequence divergence.  (d--e)~To further demonstrate that our conclusions are not a result of inadequate sequence divergence, we removed all genes with greater than 50\% mutual codon identity, as well as all loci that meet our criteria for conserved rarity (i.e., \eqsref{eq:penr} and \ref{eq:pneutral}) due to the conservation of specific codons (about 6\% of the putative conserved loci).  Then, with these ambiguous data excluded, we repeated all comparisons with our folding model.  The results shown here are consistent with Figures~3 and~4, indicating that the excluded data do not have a significant effect on our conclusions.}
  \label{fig:diversity}
\end{figure}

\begin{figure*}[h]
  \includegraphics{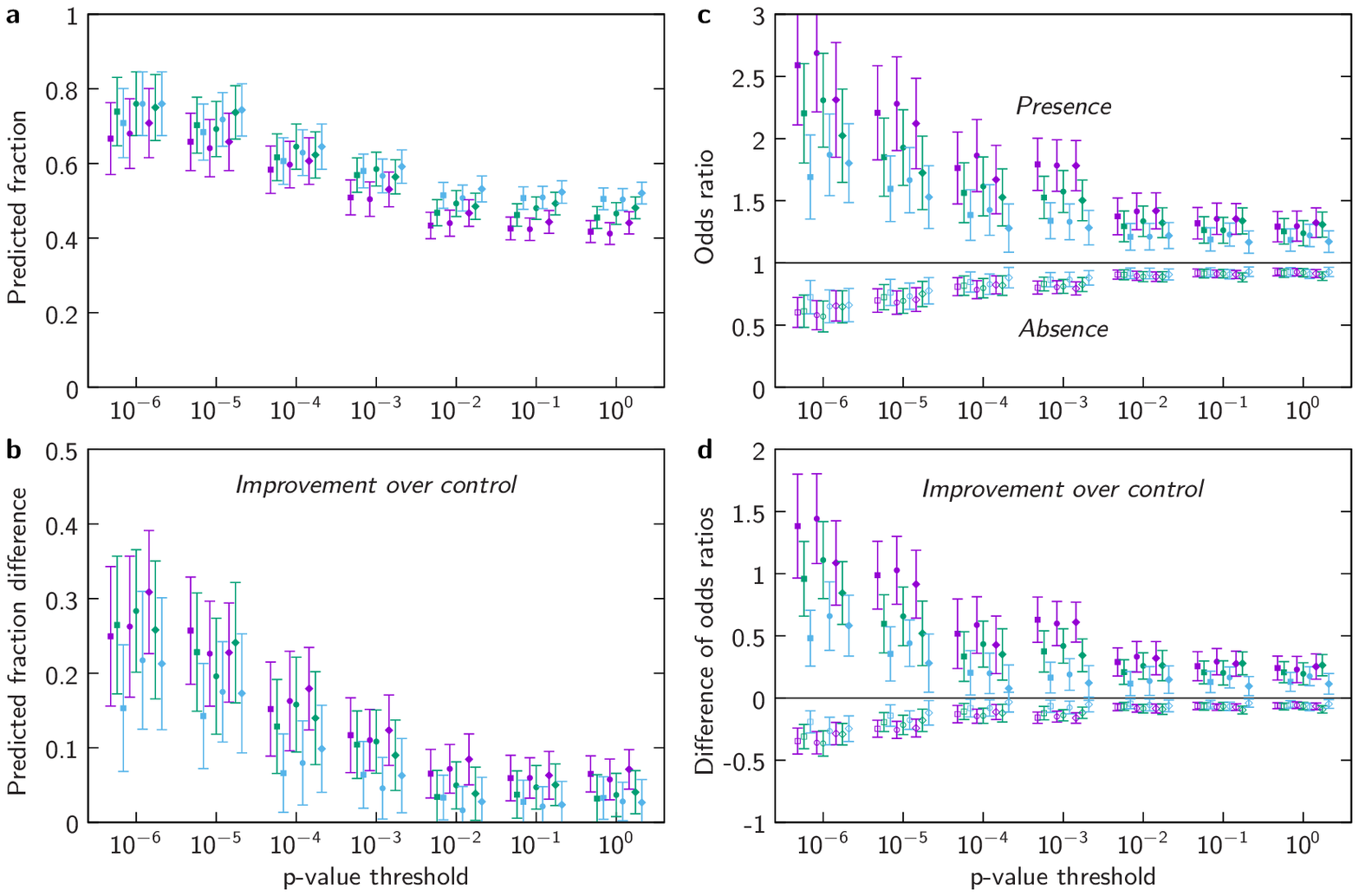}
  \caption{The co-translational folding predictions are relatively insensitive to the model parameters.  Here we vary both the consensus crystal structure definition (squares: residue--residue contacts appear in at least one crystal structure; circles: residue--residue contacts appear in at least 25\% of the crystal structures; diamonds: residue--residue contacts appear in at least 50\% of the crystal structures) and the native-state stabilities (purple: ${F(N)/N = -0.05 \kB T}$; green: ${F(N)/N = -0.075 \kB T}$; blue: ${F(N)/N = -0.1 \kB T}$, where $N$ is the protein length).  Panels (a) and (c) are analogous to Figures~3 and~4, respectively, while panels (b) and (d) show the differences between the model results and the mean of the randomized control distribution for each set of model parameters.}
  \label{fig:temperature}
\end{figure*}

\begin{figure}[h]
  \includegraphics{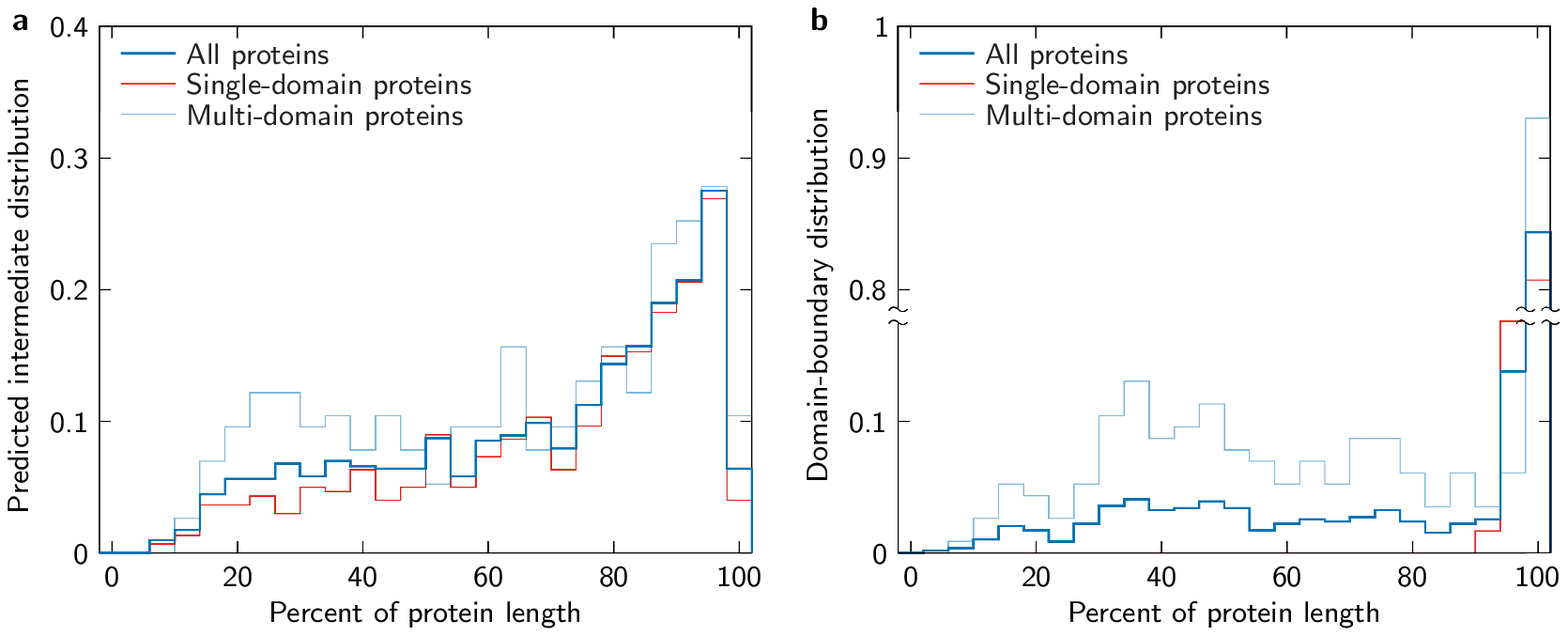}
  \caption{The distributions of (a) predicted co-translational folding intermediates and (b) domain boundaries for the data set used in this study.  The locations within the protein sequences are normalized by the protein length.  Note that some domain boundaries in single-domain proteins do not coincide with the full protein length due to small discrepancies between the crystal-structure sequences and the complete sequences extracted from the \Ecoli{} genome.}
  \label{fig:intermediate_distribution}
\end{figure}

\begin{figure}[h]
  \includegraphics{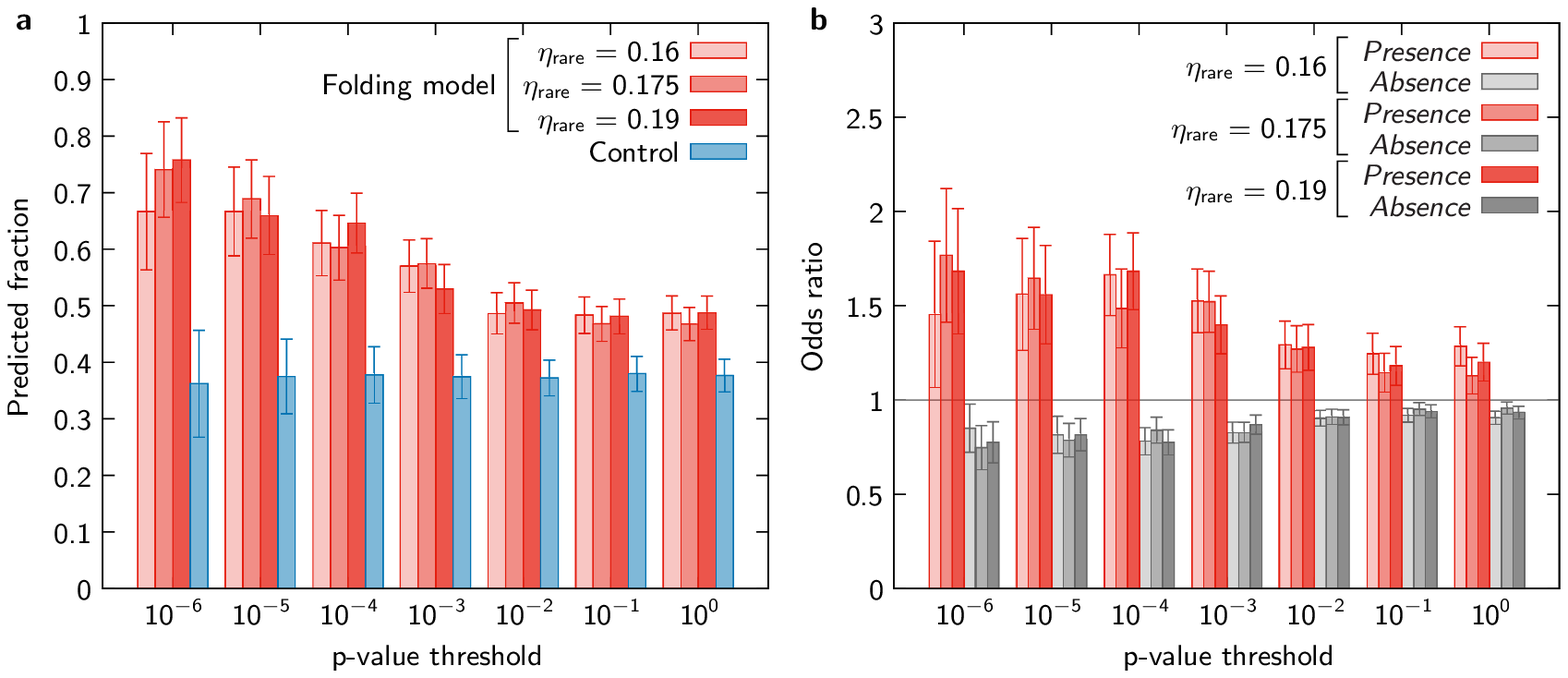}
  \caption{The conclusions of this study are robust with respect to alternative definitions of codon rarity.  Slightly different sets of rare codons can be obtained by changing the relative usage cut-off (chosen to be 10\% in the main text) or by normalizing the relative usage with respect to the usage of the most common codon for each amino acid.  Here we have changed the rare-codon definition (see \eqref{eq:pusage}) to be ${p^{s}_{\text{use}}(c | a) \le \eta_{\text{rare}} \max_a (p^{s}_{\text{use}})}$, where ${\max_a (p^{s}_{\text{use}})}$ is the relative usage of the most common codon for the amino acid~$a$ in the genome~$s$, and we have varied the cut-off ${\eta_{\text{rare}}}$ between $0.16$ and $0.19$.  We then repeated the complete calculations described in \secref{sec:neutral_model} with these rare-codon definitions.  Panels (a) and (b) show that the calculations using these alternative definitions are consistent with the results shown in Figures~3 and~4, respectively.}
  \label{fig:altrare}
\end{figure}

\begin{figure*}[h]
  \includegraphics{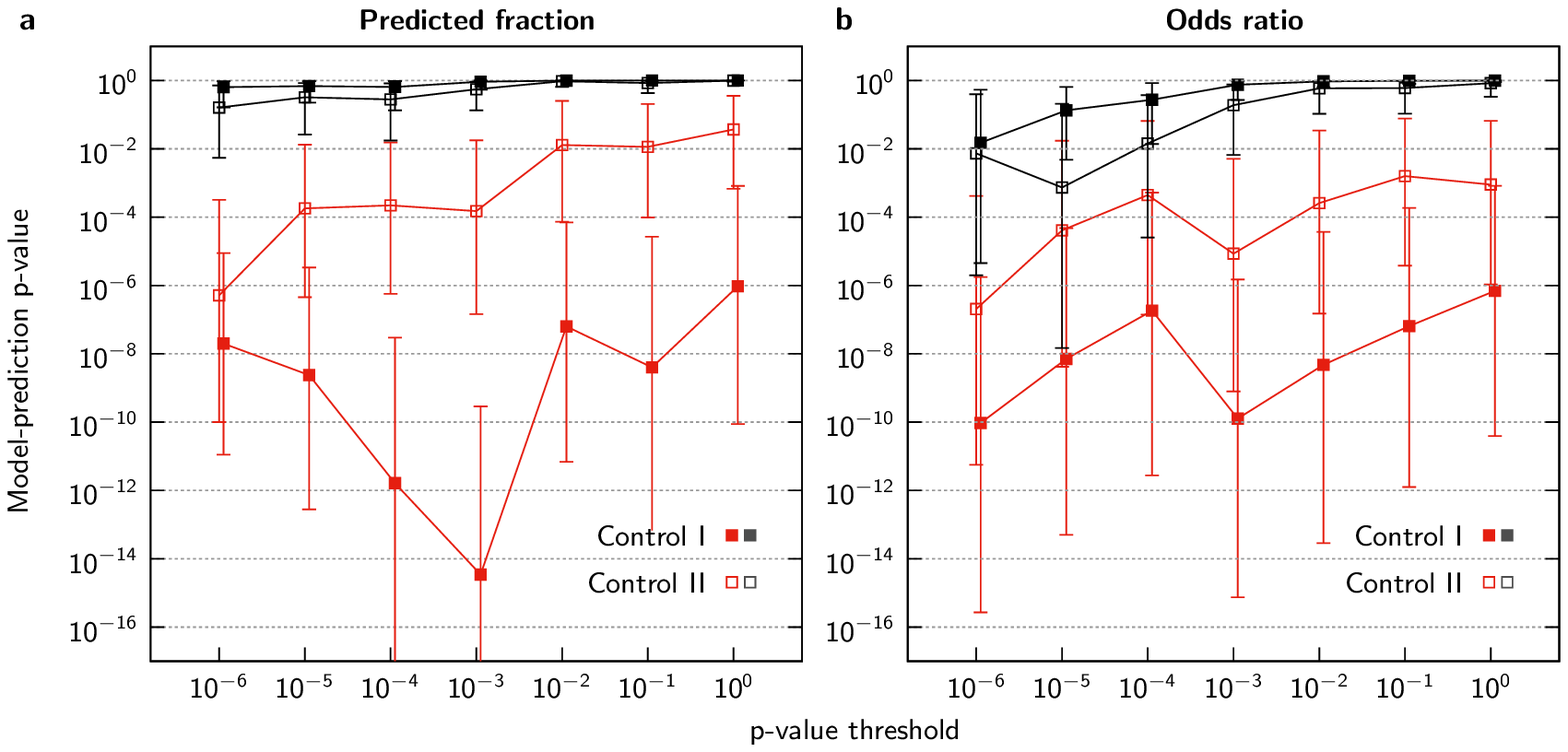}
  \caption{The probability of generating the observed correspondence between co-translational folding intermediates and putative pause sites from the randomized control data sets.  The `model-prediction p-values' are calculated from the control distributions (Figures~3, 4 and \ref{fig:alternative_control}) that were estimated from 100 independent randomizations, and should not be confused with the neutral-model p-value threshold used to identify putative pause sites.  Control I (uniform distribution of random pause sites) refers to the control discussed in the main text, while Control II (3'-end-biased pause sites) is defined in \figref{fig:alternative_control}.  Model-prediction p-values are shown in red for the co-translational folding model and in gray for the domain-boundary hypothesis.  To account for the fact that we have performed calculations using a 511-protein subset of the \Ecoli{} proteome, we computed 95\% confidence intervals (shown as error bars) for the model-prediction p-values, assuming that our subset is a representative sample of the \Ecoli{} proteome.  The overall statistical significance of our results is influenced by the number of putative pause sites, which decreases as the p-value threshold is lowered.  Consequently, the model-prediction p-values are not monotonic functions of the neutral-model p-value threshold.}
  \label{fig:pvalues}
\end{figure*}

\begin{figure*}[h]
  \includegraphics{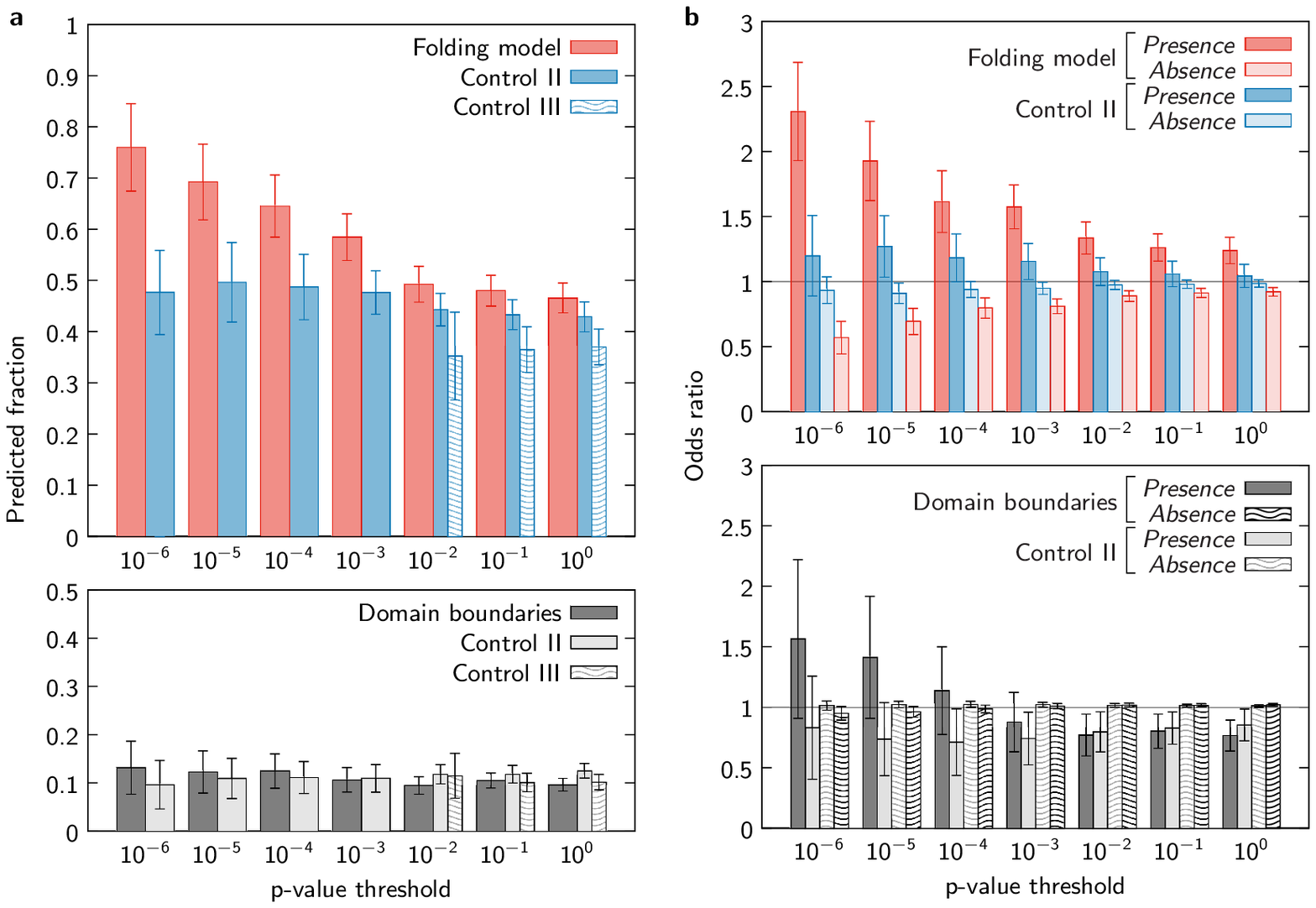}
  \caption{The performance of the co-translational folding model and the domain-boundary hypothesis relative to two alternative controls.  Panels (a) and (b) show the fraction of correctly predicted pause sites and the associated odds ratios as in Figures~3 and~4, respectively.  For Control II, randomized pause sites were generated according to the observed distribution within mRNA transcripts (\figref{fig:rc_distribution}b) at each p-value threshold, resulting in a bias towards the 3'-end.  Comparison with this control shows that the predictions of our model are not simply a result of this bias.  For Control III, reverse translations of the \Ecoli{} proteome were carried out using the gene-specific rare-codon probabilities defined in \eqsref{eq:prare}--\ref{eq:gamma}, and then all calculations were repeated as described in \secref{sec:neutral_model}.  Comparison with this control shows that our results are not a consequence of a hidden amino-acid-sequence bias.  Because the reverse translations were generated from the neutral model, we do not obtain a sufficient number of fictitious pause sites to compare with the results of our folding model at p-value thresholds less than $10^{-2}\!$.  As in Figures~3 and~4, the error bars on the controls report the standard deviation of the results obtained from 100 independent randomizations.}
  \label{fig:alternative_control}
\end{figure*}

\begin{figure*}[h]
  \includegraphics{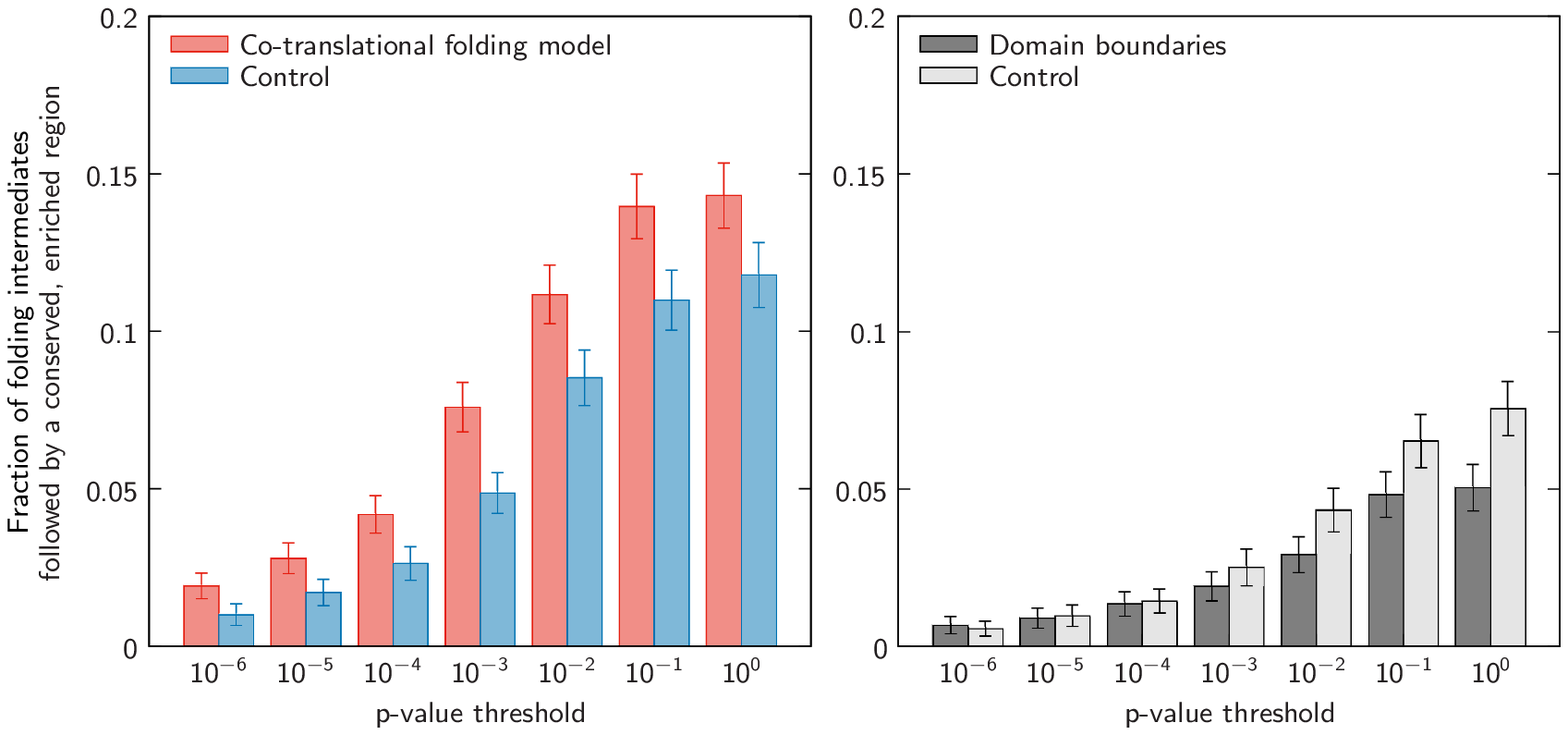}
  \caption{The fraction of predicted intermediates that are followed by conserved, rare-codon enriched regions.  On the left, the predictions of the co-translational folding model, and on the right, comparison with the domain-boundary hypothesis.  The randomized control distributions and error bars are defined as in Figure~3.  Note that, for both the folding model and the domain-boundary hypothesis, the results are strongly affected by the total number of putative conserved sites, which increases with the p-value threshold (\figref{fig:rc_distribution}a).  However, the statistical significance of the folding-model predictions is maximized at a p-value threshold of ${10^{-3}\!}$, which provides an optimal balance of false positives and false negatives in the identification of putative pause sites.}
  \label{fig:confirmed}
\end{figure*}

\begin{figure*}[h]
  \includegraphics{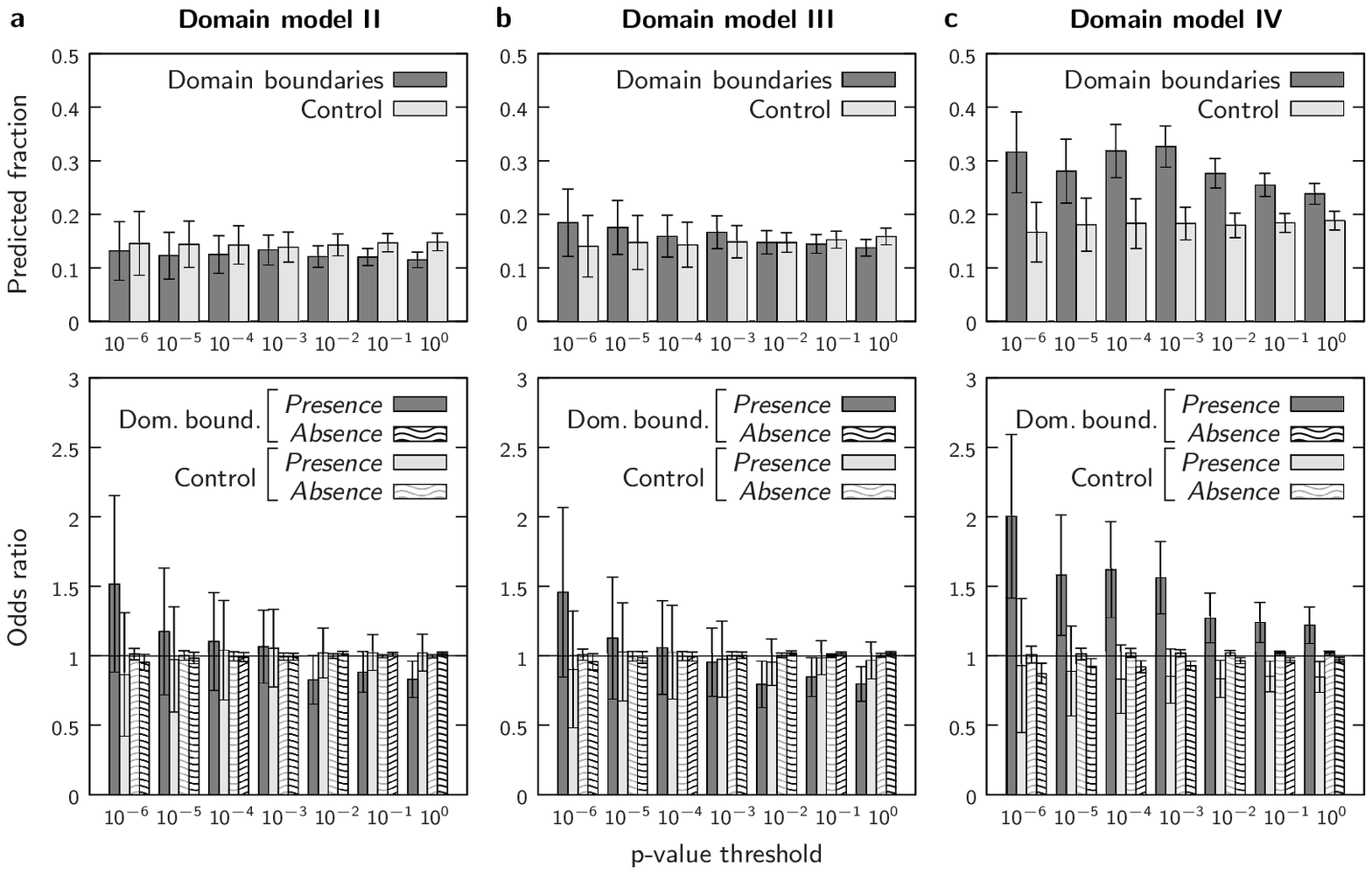}
  \caption{Alternate domain-boundary models yield similar results.  In each of these three models, we assume that the native structure forms prior to the C-terminus of the domain: model II, at 10 residues before the domain boundary; model III, at 95\% of the domain length; and model IV, at 90\% of the domain length.  Only model IV, which assumes that native structure forms well before the domain boundary for all domains, makes statistically significant predictions; nevertheless, model IV accounts for less than 35\% of the putative pause sites at all p-value thresholds.  All panels are analogous to Figure~3 (top row) or Figure~4 (bottom row).}
  \label{fig:alternative_domains}
\end{figure*}


\begin{thebibliography}{61}%
\makeatletter
\providecommand \@ifxundefined [1]{%
 \@ifx{#1\undefined}
}%
\providecommand \@ifnum [1]{%
 \ifnum #1\expandafter \@firstoftwo
 \else \expandafter \@secondoftwo
 \fi
}%
\providecommand \@ifx [1]{%
 \ifx #1\expandafter \@firstoftwo
 \else \expandafter \@secondoftwo
 \fi
}%
\providecommand \natexlab [1]{#1}%
\providecommand \enquote  [1]{``#1''}%
\providecommand \bibnamefont  [1]{#1}%
\providecommand \bibfnamefont [1]{#1}%
\providecommand \citenamefont [1]{#1}%
\providecommand \href@noop [0]{\@secondoftwo}%
\providecommand \href [0]{\begingroup \@sanitize@url \@href}%
\providecommand \@href[1]{\@@startlink{#1}\@@href}%
\providecommand \@@href[1]{\endgroup#1\@@endlink}%
\providecommand \@sanitize@url [0]{\catcode `\\12\catcode `\$12\catcode
  `\&12\catcode `\#12\catcode `\^12\catcode `\_12\catcode `\%12\relax}%
\providecommand \@@startlink[1]{}%
\providecommand \@@endlink[0]{}%
\providecommand \url  [0]{\begingroup\@sanitize@url \@url }%
\providecommand \@url [1]{\endgroup\@href {#1}{\urlprefix }}%
\providecommand \urlprefix  [0]{URL }%
\providecommand \Eprint [0]{\href }%
\providecommand \doibase [0]{http://dx.doi.org/}%
\providecommand \selectlanguage [0]{\@gobble}%
\providecommand \bibinfo  [0]{\@secondoftwo}%
\providecommand \bibfield  [0]{\@secondoftwo}%
\providecommand \translation [1]{[#1]}%
\providecommand \BibitemOpen [0]{}%
\providecommand \bibitemStop [0]{}%
\providecommand \bibitemNoStop [0]{.\EOS\space}%
\providecommand \EOS [0]{\spacefactor3000\relax}%
\providecommand \BibitemShut  [1]{\csname bibitem#1\endcsname}%
\let\auto@bib@innerbib\@empty
\bibitem [{\citenamefont {Komar}(2009)}]{komar2009pause}%
  \BibitemOpen
  \bibfield  {author} {\bibinfo {author} {\bibfnamefont {A.~A.}\ \bibnamefont
  {Komar}},\ }\href@noop {} {\bibfield  {journal} {\bibinfo  {journal} {Trends
  Biochem. Sci.}\ }\textbf {\bibinfo {volume} {34}},\ \bibinfo {pages} {16}
  (\bibinfo {year} {2009})}\BibitemShut {NoStop}%
\bibitem [{\citenamefont {Pechmann}\ \emph {et~al.}(2013)\citenamefont
  {Pechmann}, \citenamefont {Willmund},\ and\ \citenamefont
  {Frydman}}]{pechmann2013ribosome}%
  \BibitemOpen
  \bibfield  {author} {\bibinfo {author} {\bibfnamefont {S.}~\bibnamefont
  {Pechmann}}, \bibinfo {author} {\bibfnamefont {F.}~\bibnamefont {Willmund}},
  \ and\ \bibinfo {author} {\bibfnamefont {J.}~\bibnamefont {Frydman}},\
  }\href@noop {} {\bibfield  {journal} {\bibinfo  {journal} {Mol. Cell}\
  }\textbf {\bibinfo {volume} {49}},\ \bibinfo {pages} {411} (\bibinfo {year}
  {2013})}\BibitemShut {NoStop}%
\bibitem [{\citenamefont {Ciryam}\ \emph {et~al.}(2013)\citenamefont {Ciryam},
  \citenamefont {Morimoto}, \citenamefont {Vendruscolo}, \citenamefont
  {Dobson},\ and\ \citenamefont {O’Brien}}]{ciryam2013vivo}%
  \BibitemOpen
  \bibfield  {author} {\bibinfo {author} {\bibfnamefont {P.}~\bibnamefont
  {Ciryam}}, \bibinfo {author} {\bibfnamefont {R.~I.}\ \bibnamefont
  {Morimoto}}, \bibinfo {author} {\bibfnamefont {M.}~\bibnamefont
  {Vendruscolo}}, \bibinfo {author} {\bibfnamefont {C.~M.}\ \bibnamefont
  {Dobson}}, \ and\ \bibinfo {author} {\bibfnamefont {E.~P.}\ \bibnamefont
  {O’Brien}},\ }\href@noop {} {\bibfield  {journal} {\bibinfo  {journal}
  {Proc. Natl. Acad. Sci.}\ }\textbf {\bibinfo {volume} {110}},\ \bibinfo
  {pages} {E132} (\bibinfo {year} {2013})}\BibitemShut {NoStop}%
\bibitem [{\citenamefont {Netzer}\ and\ \citenamefont
  {Hartl}(1997)}]{netzer1997recombination}%
  \BibitemOpen
  \bibfield  {author} {\bibinfo {author} {\bibfnamefont {W.~J.}\ \bibnamefont
  {Netzer}}\ and\ \bibinfo {author} {\bibfnamefont {F.~U.}\ \bibnamefont
  {Hartl}},\ }\href@noop {} {\bibfield  {journal} {\bibinfo  {journal}
  {Nature}\ }\textbf {\bibinfo {volume} {388}},\ \bibinfo {pages} {343}
  (\bibinfo {year} {1997})}\BibitemShut {NoStop}%
\bibitem [{\citenamefont {Frydman}\ \emph {et~al.}(1999)\citenamefont
  {Frydman}, \citenamefont {Erdjument-Bromage}, \citenamefont {Tempst},\ and\
  \citenamefont {Hartl}}]{frydman1999cotranslational}%
  \BibitemOpen
  \bibfield  {author} {\bibinfo {author} {\bibfnamefont {J.}~\bibnamefont
  {Frydman}}, \bibinfo {author} {\bibfnamefont {H.}~\bibnamefont
  {Erdjument-Bromage}}, \bibinfo {author} {\bibfnamefont {P.}~\bibnamefont
  {Tempst}}, \ and\ \bibinfo {author} {\bibfnamefont {F.~U.}\ \bibnamefont
  {Hartl}},\ }\href@noop {} {\bibfield  {journal} {\bibinfo  {journal} {Nat.
  Struct. Mol. Biol.}\ }\textbf {\bibinfo {volume} {6}},\ \bibinfo {pages}
  {697} (\bibinfo {year} {1999})}\BibitemShut {NoStop}%
\bibitem [{\citenamefont {Komar}\ \emph {et~al.}(1999)\citenamefont {Komar},
  \citenamefont {Lesnik},\ and\ \citenamefont {Reiss}}]{komar1999synonymous}%
  \BibitemOpen
  \bibfield  {author} {\bibinfo {author} {\bibfnamefont {A.~A.}\ \bibnamefont
  {Komar}}, \bibinfo {author} {\bibfnamefont {T.}~\bibnamefont {Lesnik}}, \
  and\ \bibinfo {author} {\bibfnamefont {C.}~\bibnamefont {Reiss}},\
  }\href@noop {} {\bibfield  {journal} {\bibinfo  {journal} {FEBS Letters}\
  }\textbf {\bibinfo {volume} {462}},\ \bibinfo {pages} {387} (\bibinfo {year}
  {1999})}\BibitemShut {NoStop}%
\bibitem [{\citenamefont {Kim}\ \emph {et~al.}(2015)\citenamefont {Kim},
  \citenamefont {Yoon}, \citenamefont {Shishido}, \citenamefont {Yang},
  \citenamefont {Rooney}, \citenamefont {Barral},\ and\ \citenamefont
  {Skach}}]{kim2015translational}%
  \BibitemOpen
  \bibfield  {author} {\bibinfo {author} {\bibfnamefont {S.~J.}\ \bibnamefont
  {Kim}}, \bibinfo {author} {\bibfnamefont {J.~S.}\ \bibnamefont {Yoon}},
  \bibinfo {author} {\bibfnamefont {H.}~\bibnamefont {Shishido}}, \bibinfo
  {author} {\bibfnamefont {Z.}~\bibnamefont {Yang}}, \bibinfo {author}
  {\bibfnamefont {L.~A.}\ \bibnamefont {Rooney}}, \bibinfo {author}
  {\bibfnamefont {J.~M.}\ \bibnamefont {Barral}}, \ and\ \bibinfo {author}
  {\bibfnamefont {W.~R.}\ \bibnamefont {Skach}},\ }\href@noop {} {\bibfield
  {journal} {\bibinfo  {journal} {Science}\ }\textbf {\bibinfo {volume}
  {348}},\ \bibinfo {pages} {444} (\bibinfo {year} {2015})}\BibitemShut
  {NoStop}%
\bibitem [{\citenamefont {Siller}\ \emph {et~al.}(2010)\citenamefont {Siller},
  \citenamefont {DeZwaan}, \citenamefont {Anderson}, \citenamefont {Freeman},\
  and\ \citenamefont {Barral}}]{siller2010slowing}%
  \BibitemOpen
  \bibfield  {author} {\bibinfo {author} {\bibfnamefont {E.}~\bibnamefont
  {Siller}}, \bibinfo {author} {\bibfnamefont {D.~C.}\ \bibnamefont {DeZwaan}},
  \bibinfo {author} {\bibfnamefont {J.~F.}\ \bibnamefont {Anderson}}, \bibinfo
  {author} {\bibfnamefont {B.~C.}\ \bibnamefont {Freeman}}, \ and\ \bibinfo
  {author} {\bibfnamefont {J.~M.}\ \bibnamefont {Barral}},\ }\href@noop {}
  {\bibfield  {journal} {\bibinfo  {journal} {J. Mol. Biol.}\ }\textbf
  {\bibinfo {volume} {396}},\ \bibinfo {pages} {1310} (\bibinfo {year}
  {2010})}\BibitemShut {NoStop}%
\bibitem [{\citenamefont {Ugrinov}\ and\ \citenamefont
  {Clark}(2010)}]{ugrinov2010cotranslational}%
  \BibitemOpen
  \bibfield  {author} {\bibinfo {author} {\bibfnamefont {K.~G.}\ \bibnamefont
  {Ugrinov}}\ and\ \bibinfo {author} {\bibfnamefont {P.~L.}\ \bibnamefont
  {Clark}},\ }\href@noop {} {\bibfield  {journal} {\bibinfo  {journal}
  {Biophys. J.}\ }\textbf {\bibinfo {volume} {98}},\ \bibinfo {pages} {1312}
  (\bibinfo {year} {2010})}\BibitemShut {NoStop}%
\bibitem [{\citenamefont {Agashe}\ \emph {et~al.}(2013)\citenamefont {Agashe},
  \citenamefont {Martinez-Gomez}, \citenamefont {Drummond},\ and\ \citenamefont
  {Marx}}]{agashe2013good}%
  \BibitemOpen
  \bibfield  {author} {\bibinfo {author} {\bibfnamefont {D.}~\bibnamefont
  {Agashe}}, \bibinfo {author} {\bibfnamefont {N.~C.}\ \bibnamefont
  {Martinez-Gomez}}, \bibinfo {author} {\bibfnamefont {D.~A.}\ \bibnamefont
  {Drummond}}, \ and\ \bibinfo {author} {\bibfnamefont {C.~J.}\ \bibnamefont
  {Marx}},\ }\href@noop {} {\bibfield  {journal} {\bibinfo  {journal} {Mol.
  Biol. Evo.}\ }\textbf {\bibinfo {volume} {30}},\ \bibinfo {pages} {549}
  (\bibinfo {year} {2013})}\BibitemShut {NoStop}%
\bibitem [{\citenamefont {Clark}\ and\ \citenamefont
  {King}(2001)}]{clark2001newly}%
  \BibitemOpen
  \bibfield  {author} {\bibinfo {author} {\bibfnamefont {P.~L.}\ \bibnamefont
  {Clark}}\ and\ \bibinfo {author} {\bibfnamefont {J.}~\bibnamefont {King}},\
  }\href@noop {} {\bibfield  {journal} {\bibinfo  {journal} {J. Biol. Chem.}\
  }\textbf {\bibinfo {volume} {276}},\ \bibinfo {pages} {25411} (\bibinfo
  {year} {2001})}\BibitemShut {NoStop}%
\bibitem [{\citenamefont {Evans}\ \emph {et~al.}(2008)\citenamefont {Evans},
  \citenamefont {Sander},\ and\ \citenamefont
  {Clark}}]{evans2008cotranslational}%
  \BibitemOpen
  \bibfield  {author} {\bibinfo {author} {\bibfnamefont {M.~S.}\ \bibnamefont
  {Evans}}, \bibinfo {author} {\bibfnamefont {I.~M.}\ \bibnamefont {Sander}}, \
  and\ \bibinfo {author} {\bibfnamefont {P.~L.}\ \bibnamefont {Clark}},\
  }\href@noop {} {\bibfield  {journal} {\bibinfo  {journal} {J. Mol. Biol.}\
  }\textbf {\bibinfo {volume} {383}},\ \bibinfo {pages} {683} (\bibinfo {year}
  {2008})}\BibitemShut {NoStop}%
\bibitem [{\citenamefont {Zhang}\ \emph {et~al.}(2009)\citenamefont {Zhang},
  \citenamefont {Hubalewska},\ and\ \citenamefont
  {Ignatova}}]{zhang2009transient}%
  \BibitemOpen
  \bibfield  {author} {\bibinfo {author} {\bibfnamefont {G.}~\bibnamefont
  {Zhang}}, \bibinfo {author} {\bibfnamefont {M.}~\bibnamefont {Hubalewska}}, \
  and\ \bibinfo {author} {\bibfnamefont {Z.}~\bibnamefont {Ignatova}},\
  }\href@noop {} {\bibfield  {journal} {\bibinfo  {journal} {Nat. Struct. Mol.
  Biol.}\ }\textbf {\bibinfo {volume} {16}},\ \bibinfo {pages} {274} (\bibinfo
  {year} {2009})}\BibitemShut {NoStop}%
\bibitem [{\citenamefont {Sander}\ \emph {et~al.}(2014)\citenamefont {Sander},
  \citenamefont {Chaney},\ and\ \citenamefont {Clark}}]{sander2014expanding}%
  \BibitemOpen
  \bibfield  {author} {\bibinfo {author} {\bibfnamefont {I.~M.}\ \bibnamefont
  {Sander}}, \bibinfo {author} {\bibfnamefont {J.~L.}\ \bibnamefont {Chaney}},
  \ and\ \bibinfo {author} {\bibfnamefont {P.~L.}\ \bibnamefont {Clark}},\
  }\href@noop {} {\bibfield  {journal} {\bibinfo  {journal} {J. Am. Chem.
  Soc.}\ }\textbf {\bibinfo {volume} {136}},\ \bibinfo {pages} {858} (\bibinfo
  {year} {2014})}\BibitemShut {NoStop}%
\bibitem [{\citenamefont {Buhr}\ \emph {et~al.}(2016)\citenamefont {Buhr},
  \citenamefont {Jha}, \citenamefont {Thommen}, \citenamefont {Mittelstaet},
  \citenamefont {Kutz}, \citenamefont {Schwalbe}, \citenamefont {Rodnina},\
  and\ \citenamefont {Komar}}]{buhr2016synonymous}%
  \BibitemOpen
  \bibfield  {author} {\bibinfo {author} {\bibfnamefont {F.}~\bibnamefont
  {Buhr}}, \bibinfo {author} {\bibfnamefont {S.}~\bibnamefont {Jha}}, \bibinfo
  {author} {\bibfnamefont {M.}~\bibnamefont {Thommen}}, \bibinfo {author}
  {\bibfnamefont {J.}~\bibnamefont {Mittelstaet}}, \bibinfo {author}
  {\bibfnamefont {F.}~\bibnamefont {Kutz}}, \bibinfo {author} {\bibfnamefont
  {H.}~\bibnamefont {Schwalbe}}, \bibinfo {author} {\bibfnamefont {M.~V.}\
  \bibnamefont {Rodnina}}, \ and\ \bibinfo {author} {\bibfnamefont {A.~A.}\
  \bibnamefont {Komar}},\ }\href@noop {} {\bibfield  {journal} {\bibinfo
  {journal} {Mol. Cell}\ }\textbf {\bibinfo {volume} {61}},\ \bibinfo {pages}
  {341} (\bibinfo {year} {2016})}\BibitemShut {NoStop}%
\bibitem [{\citenamefont {Zhou}\ \emph {et~al.}(2013)\citenamefont {Zhou},
  \citenamefont {Guo}, \citenamefont {Cha}, \citenamefont {Chae}, \citenamefont
  {Chen}, \citenamefont {Barral}, \citenamefont {Sachs},\ and\ \citenamefont
  {Liu}}]{zhou2013non}%
  \BibitemOpen
  \bibfield  {author} {\bibinfo {author} {\bibfnamefont {M.}~\bibnamefont
  {Zhou}}, \bibinfo {author} {\bibfnamefont {J.}~\bibnamefont {Guo}}, \bibinfo
  {author} {\bibfnamefont {J.}~\bibnamefont {Cha}}, \bibinfo {author}
  {\bibfnamefont {M.}~\bibnamefont {Chae}}, \bibinfo {author} {\bibfnamefont
  {S.}~\bibnamefont {Chen}}, \bibinfo {author} {\bibfnamefont {J.~M.}\
  \bibnamefont {Barral}}, \bibinfo {author} {\bibfnamefont {M.~S.}\
  \bibnamefont {Sachs}}, \ and\ \bibinfo {author} {\bibfnamefont
  {Y.}~\bibnamefont {Liu}},\ }\href@noop {} {\bibfield  {journal} {\bibinfo
  {journal} {Nature}\ }\textbf {\bibinfo {volume} {495}},\ \bibinfo {pages}
  {111} (\bibinfo {year} {2013})}\BibitemShut {NoStop}%
\bibitem [{\citenamefont {O’Brien}\ \emph {et~al.}(2014)\citenamefont
  {O’Brien}, \citenamefont {Ciryam}, \citenamefont {Vendruscolo},\ and\
  \citenamefont {Dobson}}]{obrien2014understanding}%
  \BibitemOpen
  \bibfield  {author} {\bibinfo {author} {\bibfnamefont {E.~P.}\ \bibnamefont
  {O’Brien}}, \bibinfo {author} {\bibfnamefont {P.}~\bibnamefont {Ciryam}},
  \bibinfo {author} {\bibfnamefont {M.}~\bibnamefont {Vendruscolo}}, \ and\
  \bibinfo {author} {\bibfnamefont {C.~M.}\ \bibnamefont {Dobson}},\
  }\href@noop {} {\bibfield  {journal} {\bibinfo  {journal} {Acc. Chem. Res.}\
  }\textbf {\bibinfo {volume} {47}},\ \bibinfo {pages} {1536} (\bibinfo {year}
  {2014})}\BibitemShut {NoStop}%
\bibitem [{\citenamefont {Nissley}\ \emph {et~al.}(2016)\citenamefont
  {Nissley}, \citenamefont {Sharma}, \citenamefont {Ahmed}, \citenamefont
  {Friedrich}, \citenamefont {Kramer}, \citenamefont {Bukau},\ and\
  \citenamefont {O’Brien}}]{nissley2016accurate}%
  \BibitemOpen
  \bibfield  {author} {\bibinfo {author} {\bibfnamefont {D.~A.}\ \bibnamefont
  {Nissley}}, \bibinfo {author} {\bibfnamefont {A.~K.}\ \bibnamefont {Sharma}},
  \bibinfo {author} {\bibfnamefont {N.}~\bibnamefont {Ahmed}}, \bibinfo
  {author} {\bibfnamefont {U.~A.}\ \bibnamefont {Friedrich}}, \bibinfo {author}
  {\bibfnamefont {G.}~\bibnamefont {Kramer}}, \bibinfo {author} {\bibfnamefont
  {B.}~\bibnamefont {Bukau}}, \ and\ \bibinfo {author} {\bibfnamefont {E.~P.}\
  \bibnamefont {O’Brien}},\ }\href@noop {} {\bibfield  {journal} {\bibinfo
  {journal} {Nat. Comm.}\ }\textbf {\bibinfo {volume} {7}} (\bibinfo {year}
  {2016})}\BibitemShut {NoStop}%
\bibitem [{\citenamefont {Xu}\ \emph {et~al.}(2013)\citenamefont {Xu},
  \citenamefont {Ma}, \citenamefont {Shah}, \citenamefont {Rokas},
  \citenamefont {Liu},\ and\ \citenamefont {Johnson}}]{xu2013non}%
  \BibitemOpen
  \bibfield  {author} {\bibinfo {author} {\bibfnamefont {Y.}~\bibnamefont
  {Xu}}, \bibinfo {author} {\bibfnamefont {P.}~\bibnamefont {Ma}}, \bibinfo
  {author} {\bibfnamefont {P.}~\bibnamefont {Shah}}, \bibinfo {author}
  {\bibfnamefont {A.}~\bibnamefont {Rokas}}, \bibinfo {author} {\bibfnamefont
  {Y.}~\bibnamefont {Liu}}, \ and\ \bibinfo {author} {\bibfnamefont {C.~H.}\
  \bibnamefont {Johnson}},\ }\href@noop {} {\bibfield  {journal} {\bibinfo
  {journal} {Nature}\ }\textbf {\bibinfo {volume} {495}},\ \bibinfo {pages}
  {116} (\bibinfo {year} {2013})}\BibitemShut {NoStop}%
\bibitem [{\citenamefont {Kimchi-Sarfaty}\ \emph {et~al.}(2007)\citenamefont
  {Kimchi-Sarfaty}, \citenamefont {Oh}, \citenamefont {Kim}, \citenamefont
  {Sauna}, \citenamefont {Calcagno}, \citenamefont {Ambudkar},\ and\
  \citenamefont {Gottesman}}]{kimchi2007silent}%
  \BibitemOpen
  \bibfield  {author} {\bibinfo {author} {\bibfnamefont {C.}~\bibnamefont
  {Kimchi-Sarfaty}}, \bibinfo {author} {\bibfnamefont {J.~M.}\ \bibnamefont
  {Oh}}, \bibinfo {author} {\bibfnamefont {I.-W.}\ \bibnamefont {Kim}},
  \bibinfo {author} {\bibfnamefont {Z.~E.}\ \bibnamefont {Sauna}}, \bibinfo
  {author} {\bibfnamefont {A.~M.}\ \bibnamefont {Calcagno}}, \bibinfo {author}
  {\bibfnamefont {S.~V.}\ \bibnamefont {Ambudkar}}, \ and\ \bibinfo {author}
  {\bibfnamefont {M.~M.}\ \bibnamefont {Gottesman}},\ }\href@noop {} {\bibfield
   {journal} {\bibinfo  {journal} {Science}\ }\textbf {\bibinfo {volume}
  {315}},\ \bibinfo {pages} {525} (\bibinfo {year} {2007})}\BibitemShut
  {NoStop}%
\bibitem [{\citenamefont {Hartl}\ and\ \citenamefont
  {Hayer-Hartl}(2009)}]{hartl2009converging}%
  \BibitemOpen
  \bibfield  {author} {\bibinfo {author} {\bibfnamefont {F.~U.}\ \bibnamefont
  {Hartl}}\ and\ \bibinfo {author} {\bibfnamefont {M.}~\bibnamefont
  {Hayer-Hartl}},\ }\href@noop {} {\bibfield  {journal} {\bibinfo  {journal}
  {Nat. Struct. Mol. Biol.}\ }\textbf {\bibinfo {volume} {16}},\ \bibinfo
  {pages} {574} (\bibinfo {year} {2009})}\BibitemShut {NoStop}%
\bibitem [{\citenamefont {Jacobs}\ and\ \citenamefont
  {Shakhnovich}(2016)}]{jacobs2016structure}%
  \BibitemOpen
  \bibfield  {author} {\bibinfo {author} {\bibfnamefont {W.~M.}\ \bibnamefont
  {Jacobs}}\ and\ \bibinfo {author} {\bibfnamefont {E.~I.}\ \bibnamefont
  {Shakhnovich}},\ }\href@noop {} {\bibfield  {journal} {\bibinfo  {journal}
  {Biophys. J.}\ }\textbf {\bibinfo {volume} {111}},\ \bibinfo {pages} {925}
  (\bibinfo {year} {2016})}\BibitemShut {NoStop}%
\bibitem [{\citenamefont {Eichmann}\ \emph {et~al.}(2010)\citenamefont
  {Eichmann}, \citenamefont {Preissler}, \citenamefont {Riek},\ and\
  \citenamefont {Deuerling}}]{eichmann2010cotranslational}%
  \BibitemOpen
  \bibfield  {author} {\bibinfo {author} {\bibfnamefont {C.}~\bibnamefont
  {Eichmann}}, \bibinfo {author} {\bibfnamefont {S.}~\bibnamefont {Preissler}},
  \bibinfo {author} {\bibfnamefont {R.}~\bibnamefont {Riek}}, \ and\ \bibinfo
  {author} {\bibfnamefont {E.}~\bibnamefont {Deuerling}},\ }\href@noop {}
  {\bibfield  {journal} {\bibinfo  {journal} {Proc. Natl. Acad. Sci.}\ }\textbf
  {\bibinfo {volume} {107}},\ \bibinfo {pages} {9111} (\bibinfo {year}
  {2010})}\BibitemShut {NoStop}%
\bibitem [{\citenamefont {Holtkamp}\ \emph {et~al.}(2015)\citenamefont
  {Holtkamp}, \citenamefont {Kokic}, \citenamefont {J{\"a}ger}, \citenamefont
  {Mittelstaet}, \citenamefont {Komar},\ and\ \citenamefont
  {Rodnina}}]{holtkamp2015cotranslational}%
  \BibitemOpen
  \bibfield  {author} {\bibinfo {author} {\bibfnamefont {W.}~\bibnamefont
  {Holtkamp}}, \bibinfo {author} {\bibfnamefont {G.}~\bibnamefont {Kokic}},
  \bibinfo {author} {\bibfnamefont {M.}~\bibnamefont {J{\"a}ger}}, \bibinfo
  {author} {\bibfnamefont {J.}~\bibnamefont {Mittelstaet}}, \bibinfo {author}
  {\bibfnamefont {A.~A.}\ \bibnamefont {Komar}}, \ and\ \bibinfo {author}
  {\bibfnamefont {M.~V.}\ \bibnamefont {Rodnina}},\ }\href@noop {} {\bibfield
  {journal} {\bibinfo  {journal} {Science}\ }\textbf {\bibinfo {volume}
  {350}},\ \bibinfo {pages} {1104} (\bibinfo {year} {2015})}\BibitemShut
  {NoStop}%
\bibitem [{\citenamefont {Elcock}(2006)}]{elcock2006molecular}%
  \BibitemOpen
  \bibfield  {author} {\bibinfo {author} {\bibfnamefont {A.~H.}\ \bibnamefont
  {Elcock}},\ }\href@noop {} {\bibfield  {journal} {\bibinfo  {journal} {PLoS
  Comp. Biol.}\ }\textbf {\bibinfo {volume} {2}},\ \bibinfo {pages} {e98}
  (\bibinfo {year} {2006})}\BibitemShut {NoStop}%
\bibitem [{\citenamefont {O’Brien}\ \emph {et~al.}(2012)\citenamefont
  {O’Brien}, \citenamefont {Vendruscolo},\ and\ \citenamefont
  {Dobson}}]{obrien2012prediction}%
  \BibitemOpen
  \bibfield  {author} {\bibinfo {author} {\bibfnamefont {E.~P.}\ \bibnamefont
  {O’Brien}}, \bibinfo {author} {\bibfnamefont {M.}~\bibnamefont
  {Vendruscolo}}, \ and\ \bibinfo {author} {\bibfnamefont {C.~M.}\ \bibnamefont
  {Dobson}},\ }\href@noop {} {\bibfield  {journal} {\bibinfo  {journal} {Nat.
  Comm.}\ }\textbf {\bibinfo {volume} {3}},\ \bibinfo {pages} {868} (\bibinfo
  {year} {2012})}\BibitemShut {NoStop}%
\bibitem [{\citenamefont {Jacobson}\ and\ \citenamefont
  {Clark}(2016)}]{jacobson2016quality}%
  \BibitemOpen
  \bibfield  {author} {\bibinfo {author} {\bibfnamefont {G.~N.}\ \bibnamefont
  {Jacobson}}\ and\ \bibinfo {author} {\bibfnamefont {P.~L.}\ \bibnamefont
  {Clark}},\ }\href@noop {} {\bibfield  {journal} {\bibinfo  {journal} {Curr.
  Opin. Struct. Biol.}\ }\textbf {\bibinfo {volume} {38}},\ \bibinfo {pages}
  {102} (\bibinfo {year} {2016})}\BibitemShut {NoStop}%
\bibitem [{\citenamefont {Sharp}\ and\ \citenamefont
  {Li}(1987)}]{sharp1987codon}%
  \BibitemOpen
  \bibfield  {author} {\bibinfo {author} {\bibfnamefont {P.~M.}\ \bibnamefont
  {Sharp}}\ and\ \bibinfo {author} {\bibfnamefont {W.-H.}\ \bibnamefont {Li}},\
  }\href@noop {} {\bibfield  {journal} {\bibinfo  {journal} {Nucl. Acids Res.}\
  }\textbf {\bibinfo {volume} {15}},\ \bibinfo {pages} {1281} (\bibinfo {year}
  {1987})}\BibitemShut {NoStop}%
\bibitem [{\citenamefont {Dana}\ and\ \citenamefont
  {Tuller}(2014)}]{dana2014effect}%
  \BibitemOpen
  \bibfield  {author} {\bibinfo {author} {\bibfnamefont {A.}~\bibnamefont
  {Dana}}\ and\ \bibinfo {author} {\bibfnamefont {T.}~\bibnamefont {Tuller}},\
  }\href@noop {} {\bibfield  {journal} {\bibinfo  {journal} {Nucl. Acids Res.}\
  }\textbf {\bibinfo {volume} {42}},\ \bibinfo {pages} {9171} (\bibinfo {year}
  {2014})}\BibitemShut {NoStop}%
\bibitem [{\citenamefont {Li}\ \emph {et~al.}(2012)\citenamefont {Li},
  \citenamefont {Oh},\ and\ \citenamefont {Weissman}}]{li2012anti}%
  \BibitemOpen
  \bibfield  {author} {\bibinfo {author} {\bibfnamefont {G.-W.}\ \bibnamefont
  {Li}}, \bibinfo {author} {\bibfnamefont {E.}~\bibnamefont {Oh}}, \ and\
  \bibinfo {author} {\bibfnamefont {J.~S.}\ \bibnamefont {Weissman}},\
  }\href@noop {} {\bibfield  {journal} {\bibinfo  {journal} {Nature}\ }\textbf
  {\bibinfo {volume} {484}},\ \bibinfo {pages} {538} (\bibinfo {year}
  {2012})}\BibitemShut {NoStop}%
\bibitem [{\citenamefont {Pelechano}\ \emph {et~al.}(2015)\citenamefont
  {Pelechano}, \citenamefont {Wei},\ and\ \citenamefont
  {Steinmetz}}]{pelechano2015widespread}%
  \BibitemOpen
  \bibfield  {author} {\bibinfo {author} {\bibfnamefont {V.}~\bibnamefont
  {Pelechano}}, \bibinfo {author} {\bibfnamefont {W.}~\bibnamefont {Wei}}, \
  and\ \bibinfo {author} {\bibfnamefont {L.~M.}\ \bibnamefont {Steinmetz}},\
  }\href@noop {} {\bibfield  {journal} {\bibinfo  {journal} {Cell}\ }\textbf
  {\bibinfo {volume} {161}},\ \bibinfo {pages} {1400} (\bibinfo {year}
  {2015})}\BibitemShut {NoStop}%
\bibitem [{\citenamefont {Weinberg}\ \emph {et~al.}(2016)\citenamefont
  {Weinberg}, \citenamefont {Shah}, \citenamefont {Eichhorn}, \citenamefont
  {Hussmann}, \citenamefont {Plotkin},\ and\ \citenamefont
  {Bartel}}]{weinberg2016improved}%
  \BibitemOpen
  \bibfield  {author} {\bibinfo {author} {\bibfnamefont {D.~E.}\ \bibnamefont
  {Weinberg}}, \bibinfo {author} {\bibfnamefont {P.}~\bibnamefont {Shah}},
  \bibinfo {author} {\bibfnamefont {S.~W.}\ \bibnamefont {Eichhorn}}, \bibinfo
  {author} {\bibfnamefont {J.~A.}\ \bibnamefont {Hussmann}}, \bibinfo {author}
  {\bibfnamefont {J.~B.}\ \bibnamefont {Plotkin}}, \ and\ \bibinfo {author}
  {\bibfnamefont {D.~P.}\ \bibnamefont {Bartel}},\ }\href@noop {} {\bibfield
  {journal} {\bibinfo  {journal} {Cell Rep.}\ }\textbf {\bibinfo {volume}
  {14}},\ \bibinfo {pages} {1787} (\bibinfo {year} {2016})}\BibitemShut
  {NoStop}%
\bibitem [{\citenamefont {Clarke~IV}\ and\ \citenamefont
  {Clark}(2008)}]{clarke2008rare}%
  \BibitemOpen
  \bibfield  {author} {\bibinfo {author} {\bibfnamefont {T.~F.}\ \bibnamefont
  {Clarke~IV}}\ and\ \bibinfo {author} {\bibfnamefont {P.~L.}\ \bibnamefont
  {Clark}},\ }\href@noop {} {\bibfield  {journal} {\bibinfo  {journal} {PloS
  One}\ }\textbf {\bibinfo {volume} {3}},\ \bibinfo {pages} {e3412} (\bibinfo
  {year} {2008})}\BibitemShut {NoStop}%
\bibitem [{\citenamefont {Yu}\ \emph {et~al.}(2015)\citenamefont {Yu},
  \citenamefont {Dang}, \citenamefont {Zhou}, \citenamefont {Wu}, \citenamefont
  {Zhao}, \citenamefont {Sachs},\ and\ \citenamefont {Liu}}]{yu2015codon}%
  \BibitemOpen
  \bibfield  {author} {\bibinfo {author} {\bibfnamefont {C.-H.}\ \bibnamefont
  {Yu}}, \bibinfo {author} {\bibfnamefont {Y.}~\bibnamefont {Dang}}, \bibinfo
  {author} {\bibfnamefont {Z.}~\bibnamefont {Zhou}}, \bibinfo {author}
  {\bibfnamefont {C.}~\bibnamefont {Wu}}, \bibinfo {author} {\bibfnamefont
  {F.}~\bibnamefont {Zhao}}, \bibinfo {author} {\bibfnamefont {M.~S.}\
  \bibnamefont {Sachs}}, \ and\ \bibinfo {author} {\bibfnamefont
  {Y.}~\bibnamefont {Liu}},\ }\href@noop {} {\bibfield  {journal} {\bibinfo
  {journal} {Mol. Cell}\ }\textbf {\bibinfo {volume} {59}},\ \bibinfo {pages}
  {744} (\bibinfo {year} {2015})}\BibitemShut {NoStop}%
\bibitem [{\citenamefont {Chaney}\ and\ \citenamefont
  {Clark}(2015)}]{chaney2015roles}%
  \BibitemOpen
  \bibfield  {author} {\bibinfo {author} {\bibfnamefont {J.~L.}\ \bibnamefont
  {Chaney}}\ and\ \bibinfo {author} {\bibfnamefont {P.~L.}\ \bibnamefont
  {Clark}},\ }\href@noop {} {\bibfield  {journal} {\bibinfo  {journal} {Ann.
  Rev. Biophys.}\ }\textbf {\bibinfo {volume} {44}},\ \bibinfo {pages} {143}
  (\bibinfo {year} {2015})}\BibitemShut {NoStop}%
\bibitem [{\citenamefont {Spencer}\ \emph {et~al.}(2012)\citenamefont
  {Spencer}, \citenamefont {Siller}, \citenamefont {Anderson},\ and\
  \citenamefont {Barral}}]{spencer2012silent}%
  \BibitemOpen
  \bibfield  {author} {\bibinfo {author} {\bibfnamefont {P.~S.}\ \bibnamefont
  {Spencer}}, \bibinfo {author} {\bibfnamefont {E.}~\bibnamefont {Siller}},
  \bibinfo {author} {\bibfnamefont {J.~F.}\ \bibnamefont {Anderson}}, \ and\
  \bibinfo {author} {\bibfnamefont {J.~M.}\ \bibnamefont {Barral}},\
  }\href@noop {} {\bibfield  {journal} {\bibinfo  {journal} {J. Mol. Biol.}\
  }\textbf {\bibinfo {volume} {422}},\ \bibinfo {pages} {328} (\bibinfo {year}
  {2012})}\BibitemShut {NoStop}%
\bibitem [{\citenamefont {Plotkin}\ and\ \citenamefont
  {Kudla}(2011)}]{plotkin2011synonymous}%
  \BibitemOpen
  \bibfield  {author} {\bibinfo {author} {\bibfnamefont {J.~B.}\ \bibnamefont
  {Plotkin}}\ and\ \bibinfo {author} {\bibfnamefont {G.}~\bibnamefont
  {Kudla}},\ }\href@noop {} {\bibfield  {journal} {\bibinfo  {journal} {Nat.
  Rev. Genet.}\ }\textbf {\bibinfo {volume} {12}},\ \bibinfo {pages} {32}
  (\bibinfo {year} {2011})}\BibitemShut {NoStop}%
\bibitem [{\citenamefont {Pechmann}\ and\ \citenamefont
  {Frydman}(2013)}]{pechmann2013evolutionary}%
  \BibitemOpen
  \bibfield  {author} {\bibinfo {author} {\bibfnamefont {S.}~\bibnamefont
  {Pechmann}}\ and\ \bibinfo {author} {\bibfnamefont {J.}~\bibnamefont
  {Frydman}},\ }\href@noop {} {\bibfield  {journal} {\bibinfo  {journal} {Nat.
  Struct. Mol. Biol.}\ }\textbf {\bibinfo {volume} {20}},\ \bibinfo {pages}
  {237} (\bibinfo {year} {2013})}\BibitemShut {NoStop}%
\bibitem [{\citenamefont {Widmann}\ \emph {et~al.}(2008)\citenamefont
  {Widmann}, \citenamefont {Clairo}, \citenamefont {Dippon},\ and\
  \citenamefont {Pleiss}}]{widmann2008analysis}%
  \BibitemOpen
  \bibfield  {author} {\bibinfo {author} {\bibfnamefont {M.}~\bibnamefont
  {Widmann}}, \bibinfo {author} {\bibfnamefont {M.}~\bibnamefont {Clairo}},
  \bibinfo {author} {\bibfnamefont {J.}~\bibnamefont {Dippon}}, \ and\ \bibinfo
  {author} {\bibfnamefont {J.}~\bibnamefont {Pleiss}},\ }\href@noop {}
  {\bibfield  {journal} {\bibinfo  {journal} {BMC Genomics}\ }\textbf {\bibinfo
  {volume} {9}},\ \bibinfo {pages} {207} (\bibinfo {year} {2008})}\BibitemShut
  {NoStop}%
\bibitem [{\citenamefont {Clarke~IV}\ and\ \citenamefont
  {Clark}(2010)}]{clarke2010increased}%
  \BibitemOpen
  \bibfield  {author} {\bibinfo {author} {\bibfnamefont {T.~F.}\ \bibnamefont
  {Clarke~IV}}\ and\ \bibinfo {author} {\bibfnamefont {P.~L.}\ \bibnamefont
  {Clark}},\ }\href@noop {} {\bibfield  {journal} {\bibinfo  {journal} {BMC
  Genomics}\ }\textbf {\bibinfo {volume} {11}},\ \bibinfo {pages} {118}
  (\bibinfo {year} {2010})}\BibitemShut {NoStop}%
\bibitem [{\citenamefont {Trovato}\ and\ \citenamefont
  {O'Brien}(2016)}]{trovato2016insights}%
  \BibitemOpen
  \bibfield  {author} {\bibinfo {author} {\bibfnamefont {F.}~\bibnamefont
  {Trovato}}\ and\ \bibinfo {author} {\bibfnamefont {E.~P.}\ \bibnamefont
  {O'Brien}},\ }\href@noop {} {\bibfield  {journal} {\bibinfo  {journal} {Ann.
  Rev. Biophys.}\ }\textbf {\bibinfo {volume} {45}},\ \bibinfo {pages} {345}
  (\bibinfo {year} {2016})}\BibitemShut {NoStop}%
\bibitem [{\citenamefont {Ghosh}\ and\ \citenamefont
  {Dill}(2009)}]{ghosh2009computing}%
  \BibitemOpen
  \bibfield  {author} {\bibinfo {author} {\bibfnamefont {K.}~\bibnamefont
  {Ghosh}}\ and\ \bibinfo {author} {\bibfnamefont {K.~A.}\ \bibnamefont
  {Dill}},\ }\href@noop {} {\bibfield  {journal} {\bibinfo  {journal} {Proc.
  Natl. Acad. Sci.}\ }\textbf {\bibinfo {volume} {106}},\ \bibinfo {pages}
  {10649} (\bibinfo {year} {2009})}\BibitemShut {NoStop}%
\bibitem [{\citenamefont {Gloge}\ \emph {et~al.}(2014)\citenamefont {Gloge},
  \citenamefont {Becker}, \citenamefont {Kramer},\ and\ \citenamefont
  {Bukau}}]{gloge2014co}%
  \BibitemOpen
  \bibfield  {author} {\bibinfo {author} {\bibfnamefont {F.}~\bibnamefont
  {Gloge}}, \bibinfo {author} {\bibfnamefont {A.~H.}\ \bibnamefont {Becker}},
  \bibinfo {author} {\bibfnamefont {G.}~\bibnamefont {Kramer}}, \ and\ \bibinfo
  {author} {\bibfnamefont {B.}~\bibnamefont {Bukau}},\ }\href@noop {}
  {\bibfield  {journal} {\bibinfo  {journal} {Curr. Opin. Struct. Biol.}\
  }\textbf {\bibinfo {volume} {24}},\ \bibinfo {pages} {24} (\bibinfo {year}
  {2014})}\BibitemShut {NoStop}%
\bibitem [{\citenamefont {Lu}\ and\ \citenamefont
  {Deutsch}(2005)}]{lu2005folding}%
  \BibitemOpen
  \bibfield  {author} {\bibinfo {author} {\bibfnamefont {J.}~\bibnamefont
  {Lu}}\ and\ \bibinfo {author} {\bibfnamefont {C.}~\bibnamefont {Deutsch}},\
  }\href@noop {} {\bibfield  {journal} {\bibinfo  {journal} {Nat. Struct. Mol.
  Biol.}\ }\textbf {\bibinfo {volume} {12}},\ \bibinfo {pages} {1123} (\bibinfo
  {year} {2005})}\BibitemShut {NoStop}%
\bibitem [{\citenamefont {Nilsson}\ \emph {et~al.}(2015)\citenamefont
  {Nilsson}, \citenamefont {Hedman}, \citenamefont {Marino}, \citenamefont
  {Wickles}, \citenamefont {Bischoff}, \citenamefont {Johansson}, \citenamefont
  {M{\"u}ller-Lucks}, \citenamefont {Trovato}, \citenamefont {Puglisi},
  \citenamefont {O’Brien}, \citenamefont {Beckmann},\ and\ \citenamefont {con
  Heijne}}]{nilsson2015cotranslational}%
  \BibitemOpen
  \bibfield  {author} {\bibinfo {author} {\bibfnamefont {O.~B.}\ \bibnamefont
  {Nilsson}}, \bibinfo {author} {\bibfnamefont {R.}~\bibnamefont {Hedman}},
  \bibinfo {author} {\bibfnamefont {J.}~\bibnamefont {Marino}}, \bibinfo
  {author} {\bibfnamefont {S.}~\bibnamefont {Wickles}}, \bibinfo {author}
  {\bibfnamefont {L.}~\bibnamefont {Bischoff}}, \bibinfo {author}
  {\bibfnamefont {M.}~\bibnamefont {Johansson}}, \bibinfo {author}
  {\bibfnamefont {A.}~\bibnamefont {M{\"u}ller-Lucks}}, \bibinfo {author}
  {\bibfnamefont {F.}~\bibnamefont {Trovato}}, \bibinfo {author} {\bibfnamefont
  {J.~D.}\ \bibnamefont {Puglisi}}, \bibinfo {author} {\bibfnamefont {E.~P.}\
  \bibnamefont {O’Brien}}, \bibinfo {author} {\bibfnamefont {R.}~\bibnamefont
  {Beckmann}}, \ and\ \bibinfo {author} {\bibfnamefont {G.}~\bibnamefont {con
  Heijne}},\ }\href@noop {} {\bibfield  {journal} {\bibinfo  {journal} {Cell
  Rep.}\ }\textbf {\bibinfo {volume} {12}},\ \bibinfo {pages} {1533} (\bibinfo
  {year} {2015})}\BibitemShut {NoStop}%
\bibitem [{\citenamefont {Purvis}\ \emph {et~al.}(1987)\citenamefont {Purvis},
  \citenamefont {Bettany}, \citenamefont {Santiago}, \citenamefont {Coggins},
  \citenamefont {Duncan}, \citenamefont {Eason},\ and\ \citenamefont
  {Brown}}]{purvis1987efficiency}%
  \BibitemOpen
  \bibfield  {author} {\bibinfo {author} {\bibfnamefont {I.~J.}\ \bibnamefont
  {Purvis}}, \bibinfo {author} {\bibfnamefont {A.~J.}\ \bibnamefont {Bettany}},
  \bibinfo {author} {\bibfnamefont {T.~C.}\ \bibnamefont {Santiago}}, \bibinfo
  {author} {\bibfnamefont {J.~R.}\ \bibnamefont {Coggins}}, \bibinfo {author}
  {\bibfnamefont {K.}~\bibnamefont {Duncan}}, \bibinfo {author} {\bibfnamefont
  {R.}~\bibnamefont {Eason}}, \ and\ \bibinfo {author} {\bibfnamefont {A.~J.}\
  \bibnamefont {Brown}},\ }\href@noop {} {\bibfield  {journal} {\bibinfo
  {journal} {J. Mol. Biol.}\ }\textbf {\bibinfo {volume} {193}},\ \bibinfo
  {pages} {413} (\bibinfo {year} {1987})}\BibitemShut {NoStop}%
\bibitem [{\citenamefont {Murzin}\ \emph {et~al.}(1995)\citenamefont {Murzin},
  \citenamefont {Brenner}, \citenamefont {Hubbard},\ and\ \citenamefont
  {Chothia}}]{murzin1995scop}%
  \BibitemOpen
  \bibfield  {author} {\bibinfo {author} {\bibfnamefont {A.~G.}\ \bibnamefont
  {Murzin}}, \bibinfo {author} {\bibfnamefont {S.~E.}\ \bibnamefont {Brenner}},
  \bibinfo {author} {\bibfnamefont {T.}~\bibnamefont {Hubbard}}, \ and\
  \bibinfo {author} {\bibfnamefont {C.}~\bibnamefont {Chothia}},\ }\href@noop
  {} {\bibfield  {journal} {\bibinfo  {journal} {J. Mol. Biol.}\ }\textbf
  {\bibinfo {volume} {247}},\ \bibinfo {pages} {536} (\bibinfo {year}
  {1995})}\BibitemShut {NoStop}%
\bibitem [{\citenamefont {Sharma}\ \emph {et~al.}(2016)\citenamefont {Sharma},
  \citenamefont {Bukau},\ and\ \citenamefont {O’Brien}}]{sharma2016physical}%
  \BibitemOpen
  \bibfield  {author} {\bibinfo {author} {\bibfnamefont {A.~K.}\ \bibnamefont
  {Sharma}}, \bibinfo {author} {\bibfnamefont {B.}~\bibnamefont {Bukau}}, \
  and\ \bibinfo {author} {\bibfnamefont {E.~P.}\ \bibnamefont {O’Brien}},\
  }\href@noop {} {\bibfield  {journal} {\bibinfo  {journal} {J. Am. Chem.
  Soc.}\ }\textbf {\bibinfo {volume} {138}},\ \bibinfo {pages} {1180} (\bibinfo
  {year} {2016})}\BibitemShut {NoStop}%
\bibitem [{\citenamefont {Nilsson}\ \emph {et~al.}(2017)\citenamefont
  {Nilsson}, \citenamefont {Nickson}, \citenamefont {Hollins}, \citenamefont
  {Wickles}, \citenamefont {Steward}, \citenamefont {Beckmann}, \citenamefont
  {von Heijne},\ and\ \citenamefont {Clarke}}]{nilsson2017cotranslational}%
  \BibitemOpen
  \bibfield  {author} {\bibinfo {author} {\bibfnamefont {O.~B.}\ \bibnamefont
  {Nilsson}}, \bibinfo {author} {\bibfnamefont {A.~A.}\ \bibnamefont
  {Nickson}}, \bibinfo {author} {\bibfnamefont {J.~J.}\ \bibnamefont
  {Hollins}}, \bibinfo {author} {\bibfnamefont {S.}~\bibnamefont {Wickles}},
  \bibinfo {author} {\bibfnamefont {A.}~\bibnamefont {Steward}}, \bibinfo
  {author} {\bibfnamefont {R.}~\bibnamefont {Beckmann}}, \bibinfo {author}
  {\bibfnamefont {G.}~\bibnamefont {von Heijne}}, \ and\ \bibinfo {author}
  {\bibfnamefont {J.}~\bibnamefont {Clarke}},\ }\href@noop {} {\bibfield
  {journal} {\bibinfo  {journal} {Nat. Struct. Mol. Biol.}\ }\textbf {\bibinfo
  {volume} {24}},\ \bibinfo {pages} {221} (\bibinfo {year} {2017})}\BibitemShut
  {NoStop}%
\bibitem [{\citenamefont {O'Brien}\ \emph {et~al.}(2014)\citenamefont
  {O'Brien}, \citenamefont {Vendruscolo},\ and\ \citenamefont
  {Dobson}}]{obrien2014kinetic}%
  \BibitemOpen
  \bibfield  {author} {\bibinfo {author} {\bibfnamefont {E.~P.}\ \bibnamefont
  {O'Brien}}, \bibinfo {author} {\bibfnamefont {M.}~\bibnamefont
  {Vendruscolo}}, \ and\ \bibinfo {author} {\bibfnamefont {C.~M.}\ \bibnamefont
  {Dobson}},\ }\href@noop {} {\bibfield  {journal} {\bibinfo  {journal} {Nat.
  Comm.}\ }\textbf {\bibinfo {volume} {5}},\ \bibinfo {pages} {2988} (\bibinfo
  {year} {2014})}\BibitemShut {NoStop}%
\bibitem [{\citenamefont {Trovato}\ and\ \citenamefont
  {O’Brien}(2017)}]{trovato2017fast}%
  \BibitemOpen
  \bibfield  {author} {\bibinfo {author} {\bibfnamefont {F.}~\bibnamefont
  {Trovato}}\ and\ \bibinfo {author} {\bibfnamefont {E.~P.}\ \bibnamefont
  {O’Brien}},\ }\href@noop {} {\bibfield  {journal} {\bibinfo  {journal}
  {Biophys. J.}\ }\textbf {\bibinfo {volume} {112}},\ \bibinfo {pages} {1807}
  (\bibinfo {year} {2017})}\BibitemShut {NoStop}%
\bibitem [{\citenamefont {Thanaraj}\ and\ \citenamefont
  {Argos}(1996)}]{thanaraj1996ribosome}%
  \BibitemOpen
  \bibfield  {author} {\bibinfo {author} {\bibfnamefont {T.~A.}\ \bibnamefont
  {Thanaraj}}\ and\ \bibinfo {author} {\bibfnamefont {P.}~\bibnamefont
  {Argos}},\ }\href@noop {} {\bibfield  {journal} {\bibinfo  {journal} {Protein
  Sci.}\ }\textbf {\bibinfo {volume} {5}},\ \bibinfo {pages} {1594} (\bibinfo
  {year} {1996})}\BibitemShut {NoStop}%
\bibitem [{\citenamefont {Lee}\ \emph {et~al.}(2010)\citenamefont {Lee},
  \citenamefont {Zhou}, \citenamefont {Tartaglia}, \citenamefont
  {Vendruscolo},\ and\ \citenamefont {Wilke}}]{lee2010translationally}%
  \BibitemOpen
  \bibfield  {author} {\bibinfo {author} {\bibfnamefont {Y.}~\bibnamefont
  {Lee}}, \bibinfo {author} {\bibfnamefont {T.}~\bibnamefont {Zhou}}, \bibinfo
  {author} {\bibfnamefont {G.~G.}\ \bibnamefont {Tartaglia}}, \bibinfo {author}
  {\bibfnamefont {M.}~\bibnamefont {Vendruscolo}}, \ and\ \bibinfo {author}
  {\bibfnamefont {C.~O.}\ \bibnamefont {Wilke}},\ }\href@noop {} {\bibfield
  {journal} {\bibinfo  {journal} {Proteomics}\ }\textbf {\bibinfo {volume}
  {10}},\ \bibinfo {pages} {4163} (\bibinfo {year} {2010})}\BibitemShut
  {NoStop}%
\bibitem [{\citenamefont {Chartier}\ \emph {et~al.}(2012)\citenamefont
  {Chartier}, \citenamefont {Gaudreault},\ and\ \citenamefont
  {Najmanovich}}]{chartier2012large}%
  \BibitemOpen
  \bibfield  {author} {\bibinfo {author} {\bibfnamefont {M.}~\bibnamefont
  {Chartier}}, \bibinfo {author} {\bibfnamefont {F.}~\bibnamefont
  {Gaudreault}}, \ and\ \bibinfo {author} {\bibfnamefont {R.}~\bibnamefont
  {Najmanovich}},\ }\href@noop {} {\bibfield  {journal} {\bibinfo  {journal}
  {Bioinformatics}\ }\textbf {\bibinfo {volume} {28}},\ \bibinfo {pages} {1438}
  (\bibinfo {year} {2012})}\BibitemShut {NoStop}%
\bibitem [{\citenamefont {Brunak}\ and\ \citenamefont
  {Engelbrecht}(1996)}]{brunak1996protein}%
  \BibitemOpen
  \bibfield  {author} {\bibinfo {author} {\bibfnamefont {S.}~\bibnamefont
  {Brunak}}\ and\ \bibinfo {author} {\bibfnamefont {J.}~\bibnamefont
  {Engelbrecht}},\ }\href@noop {} {\bibfield  {journal} {\bibinfo  {journal}
  {Proteins}\ }\textbf {\bibinfo {volume} {25}},\ \bibinfo {pages} {237}
  (\bibinfo {year} {1996})}\BibitemShut {NoStop}%
\bibitem [{\citenamefont {Zhou}\ \emph {et~al.}(2009)\citenamefont {Zhou},
  \citenamefont {Weems},\ and\ \citenamefont
  {Wilke}}]{zhou2009translationally}%
  \BibitemOpen
  \bibfield  {author} {\bibinfo {author} {\bibfnamefont {T.}~\bibnamefont
  {Zhou}}, \bibinfo {author} {\bibfnamefont {M.}~\bibnamefont {Weems}}, \ and\
  \bibinfo {author} {\bibfnamefont {C.~O.}\ \bibnamefont {Wilke}},\ }\href@noop
  {} {\bibfield  {journal} {\bibinfo  {journal} {Mol. Biol. Evo.}\ }\textbf
  {\bibinfo {volume} {26}},\ \bibinfo {pages} {1571} (\bibinfo {year}
  {2009})}\BibitemShut {NoStop}%
\bibitem [{\citenamefont {Chaney}\ \emph {et~al.}(2017)\citenamefont {Chaney},
  \citenamefont {Steele}, \citenamefont {Carmichael}, \citenamefont
  {Rodriguez}, \citenamefont {Specht}, \citenamefont {Ngo}, \citenamefont {Li},
  \citenamefont {Emrich},\ and\ \citenamefont {Clark}}]{chaney2017widespread}%
  \BibitemOpen
  \bibfield  {author} {\bibinfo {author} {\bibfnamefont {J.~L.}\ \bibnamefont
  {Chaney}}, \bibinfo {author} {\bibfnamefont {A.}~\bibnamefont {Steele}},
  \bibinfo {author} {\bibfnamefont {R.}~\bibnamefont {Carmichael}}, \bibinfo
  {author} {\bibfnamefont {A.}~\bibnamefont {Rodriguez}}, \bibinfo {author}
  {\bibfnamefont {A.~T.}\ \bibnamefont {Specht}}, \bibinfo {author}
  {\bibfnamefont {K.}~\bibnamefont {Ngo}}, \bibinfo {author} {\bibfnamefont
  {J.}~\bibnamefont {Li}}, \bibinfo {author} {\bibfnamefont {S.}~\bibnamefont
  {Emrich}}, \ and\ \bibinfo {author} {\bibfnamefont {P.~L.}\ \bibnamefont
  {Clark}},\ }\href@noop {} {\bibfield  {journal} {\bibinfo  {journal} {PLoS
  Comp. Biol.}\ }\textbf {\bibinfo {volume} {13}},\ \bibinfo {pages} {e1005531}
  (\bibinfo {year} {2017})}\BibitemShut {NoStop}%
\bibitem [{\citenamefont {Kudla}\ \emph {et~al.}(2009)\citenamefont {Kudla},
  \citenamefont {Murray}, \citenamefont {Tollervey},\ and\ \citenamefont
  {Plotkin}}]{kudla2009coding}%
  \BibitemOpen
  \bibfield  {author} {\bibinfo {author} {\bibfnamefont {G.}~\bibnamefont
  {Kudla}}, \bibinfo {author} {\bibfnamefont {A.~W.}\ \bibnamefont {Murray}},
  \bibinfo {author} {\bibfnamefont {D.}~\bibnamefont {Tollervey}}, \ and\
  \bibinfo {author} {\bibfnamefont {J.~B.}\ \bibnamefont {Plotkin}},\
  }\href@noop {} {\bibfield  {journal} {\bibinfo  {journal} {Science}\ }\textbf
  {\bibinfo {volume} {324}},\ \bibinfo {pages} {255} (\bibinfo {year}
  {2009})}\BibitemShut {NoStop}%
\bibitem [{\citenamefont {Goodman}\ \emph {et~al.}(2013)\citenamefont
  {Goodman}, \citenamefont {Church},\ and\ \citenamefont
  {Kosuri}}]{goodman2013causes}%
  \BibitemOpen
  \bibfield  {author} {\bibinfo {author} {\bibfnamefont {D.~B.}\ \bibnamefont
  {Goodman}}, \bibinfo {author} {\bibfnamefont {G.~M.}\ \bibnamefont {Church}},
  \ and\ \bibinfo {author} {\bibfnamefont {S.}~\bibnamefont {Kosuri}},\
  }\href@noop {} {\bibfield  {journal} {\bibinfo  {journal} {Science}\ }\textbf
  {\bibinfo {volume} {342}},\ \bibinfo {pages} {475} (\bibinfo {year}
  {2013})}\BibitemShut {NoStop}%
\bibitem [{\citenamefont {Berman}\ \emph {et~al.}(2000)\citenamefont {Berman},
  \citenamefont {Westbrook}, \citenamefont {Feng}, \citenamefont {Gilliland},
  \citenamefont {Bhat}, \citenamefont {Weissig}, \citenamefont {Shindyalov},\
  and\ \citenamefont {Bourne}}]{berman2000protein}%
  \BibitemOpen
  \bibfield  {author} {\bibinfo {author} {\bibfnamefont {H.~M.}\ \bibnamefont
  {Berman}}, \bibinfo {author} {\bibfnamefont {J.}~\bibnamefont {Westbrook}},
  \bibinfo {author} {\bibfnamefont {Z.}~\bibnamefont {Feng}}, \bibinfo {author}
  {\bibfnamefont {G.}~\bibnamefont {Gilliland}}, \bibinfo {author}
  {\bibfnamefont {T.~N.}\ \bibnamefont {Bhat}}, \bibinfo {author}
  {\bibfnamefont {H.}~\bibnamefont {Weissig}}, \bibinfo {author} {\bibfnamefont
  {I.~N.}\ \bibnamefont {Shindyalov}}, \ and\ \bibinfo {author} {\bibfnamefont
  {P.~E.}\ \bibnamefont {Bourne}},\ }\href@noop {} {\bibfield  {journal}
  {\bibinfo  {journal} {Nucl. Acids Res.}\ }\textbf {\bibinfo {volume} {28}},\
  \bibinfo {pages} {235} (\bibinfo {year} {2000})}\BibitemShut {NoStop}%
\bibitem [{\citenamefont {Wang}\ \emph {et~al.}(2012)\citenamefont {Wang},
  \citenamefont {Weiss}, \citenamefont {Simonovic}, \citenamefont {Haertinger},
  \citenamefont {Schrimpf}, \citenamefont {Hengartner},\ and\ \citenamefont
  {von Mering}}]{wang2012paxdb}%
  \BibitemOpen
  \bibfield  {author} {\bibinfo {author} {\bibfnamefont {M.}~\bibnamefont
  {Wang}}, \bibinfo {author} {\bibfnamefont {M.}~\bibnamefont {Weiss}},
  \bibinfo {author} {\bibfnamefont {M.}~\bibnamefont {Simonovic}}, \bibinfo
  {author} {\bibfnamefont {G.}~\bibnamefont {Haertinger}}, \bibinfo {author}
  {\bibfnamefont {S.~P.}\ \bibnamefont {Schrimpf}}, \bibinfo {author}
  {\bibfnamefont {M.~O.}\ \bibnamefont {Hengartner}}, \ and\ \bibinfo {author}
  {\bibfnamefont {C.}~\bibnamefont {von Mering}},\ }\href@noop {} {\bibfield
  {journal} {\bibinfo  {journal} {Mol. Cell. Proteomics}\ }\textbf {\bibinfo
  {volume} {11}},\ \bibinfo {pages} {492} (\bibinfo {year} {2012})}\BibitemShut
  {NoStop}%
\end{thebibliography}
\end{document}